# *Crystallization of Binary Nanocrystal Superlattices and the Relevance of Short-Range Attraction*

*Emanuele Marino,[1,8,‡] R. Allen LaCour,[4,‡] Timothy C. Moore,[4] Sjoerd W. van Dongen,[1,6] Austin W. Keller,[2] Di An,[1] Shengsong Yang,[1] Daniel J. Rosen,[2] Guillaume Gouget,[1] Esther H.R. Tsai,[7] Cherie R. Kagan,[1,2,3] Thomas E. Kodger,[6] Sharon C. Glotzer,[4,5,*] and Christopher B. Murray[1,2,*]*

[1]Department of Chemistry, [2]Department of Materials Science and Engineering, [3]Department of Electrical and Systems Engineering, University of Pennsylvania. Philadelphia, Pennsylvania, 19104, United States.

[4]Department of Chemical Engineering and [5]Biointerfaces Institute, University of Michigan, Ann Arbor, Michigan 48109, United States.

[6]Physical Chemistry and Soft Matter, Wageningen University and Research, 6708WE, Wageningen, The Netherlands.

[7]Brookhaven National Laboratory, Bldg. 735, Center for Functional Nanomaterials, Upton, NY 11973-5000, USA.

[8]Dipartimento di Fisica e Chimica, Università degli Studi di Palermo, Via Archirafi 36, 90123 Palermo, Italy.

ABSTRACT

The synthesis of binary nanocrystal superlattices (BNSLs) enables the targeted integration of orthogonal physical properties, like photoluminescence and magnetism, into a single superstructure, unlocking a vast design space for multifunctional materials. Yet, the formation mechanism of BNSLs remains poorly understood, restricting the use of simulation to predict the structure and properties of the superlattices. Here, we use a combination of *in situ* scattering and molecular simulation to elucidate the self-assembly of two common BNSLs through emulsion templating. Our self-assembly experiments reveal that no intermediate structures precede the formation of the final binary phases, indicating that their formation proceeds through classical nucleation. Using simulations, we find that, despite the formation of $AlB_2$ and $NaZn_{13}$ typically being attributed to entropy, their self-assembly is most consistent with the nanocrystals possessing short-range interparticle attraction, which we find can dramatically accelerate nucleation kinetics in BNSLs. We also find homogenous, classical nucleation in simulation, corroborating our experiments. These results establish a robust correspondence between experiment and theory, paving the way towards *a priori* prediction of BNSLs.

**Introduction**

Recent advances have enabled the synthesis of colloidal nanocrystals (NCs) with different sizes, shapes, and compositions, generating a library of nanoscale building blocks with well-defined optical, electronic, and magnetic properties. These properties have been exploited to develop optoelectronic devices like photodetectors,[1,2] light-emitting diodes,[3,4] field-effect transistors,[5,6] and solar cells[7,8] by assembling NCs into ordered solids, or superlattices. While single-component NC superlattices have already revealed structure-property relationships,[9,10] multi-component NC superlattices are still in the early stages of investigation.[11] The integration of NCs with orthogonal functionalities is crucial in unlocking a vast synthetic design space for material properties resulting from the synergistic interaction of the individual components.[12-20] So far, the exploration of this design space has been restricted by our limited understanding of the formation of multi-component NC superlattices.

Binary nanocrystal superlattices (BNSLs) with diverse crystal structures have been synthesized, integrating combinations of semiconducting, magnetic, and metallic NCs.[21-23] However, predicting which binary structure self-assembles from a given combination of NCs has proven extremely challenging.[24] With rare exceptions,[25-27] simulation models of binary mixtures of NCs frequently fail to self-assemble, indicating that the current understanding of how NC interactions contribute to the synthesis of BNSLs is incomplete. By contrast, experimental *in situ* studies have already revealed the self-assembly mechanism of single-component NC superlattices,[28-36] enabling *a priori* structure prediction by capturing both kinetic and thermodynamic aspects of how different inter-NC interactions influence crystallization.[37-43] Yet, in almost two decades since the first observation of BNSLs,[21] only one *in situ* study has focused on mono-functional BNSLs,[44] while no *in situ* studies have investigated the synthesis of multi-functional BNLSs.

Here we combine experiments and simulations to understand the synthesis of multi-functional BNSLs from a combination of magnetic, semiconducting, and plasmonic NCs. A combination of magnetic and semiconducting NCs may prove crucial in modulating the temperature of the atomic lattice and the conductivity of charge carriers.[45,46] Instead, a combination of magnetic and plasmonic NCs may allow active control over the coherent oscillations of charge carriers and the optical response of the material.[47,48] We use synchrotron-based *in situ* small-angle X-ray scattering (SAXS) to follow in real-time the self-assembly of BNSLs isostructural to $AlB_2$ and $NaZn_{13}$. The NCs were confined to emulsion droplets that were slowly dried to trigger crystallization, resolving with unprecedented detail the formation of high-quality BNSLs.[28,49-51] This approach was combined with molecular dynamics (MD) simulations to determine the interparticle interactions allowing the formation of these BNSLs. In contrast to some reports of single-component superlattices,[28,30,31,52] we find that nucleation occurs classically for both superlattices, although significant structural compression occurs after nucleation, indicating that the ligands are still swollen with solvent at the time of nucleation. In our MD simulations, we find that, despite the self-assembly of $AlB_2$ and $NaZn_{13}$ typically being attributed to entropy, only NCs interacting with a short-ranged attraction between NCs results in self-assembly behavior consistent with experiment. Specifically, we find that short-range attraction can dramatically accelerate the crystallizations kinetics of BNSLs. By establishing a direct link between experiments and simulations, our work provides crucial insights into the synthesis of BNSLs, providing a significant step towards the targeted synthesis and *a priori* structure prediction of these complex, 3D artificial solids.

**Results and Discussion**

We synthesize BNSLs by the self-assembly of binary mixtures of NCs using emulsion-templating,[49,53] which has recently emerged as a consistent and scalable method to fabricate 3D single-component NC superlattices.[28,29,53-61] We prepare a surfactant-stabilized oil-in-water emulsion containing a dispersion of larger (L) super-paramagnetic $Fe_3O_4$ and smaller (S) semiconductor PbS NCs with an effective size ratio of 0.56, a number ratio of 1:2, and a total inorganic volume fraction of $\phi = 0.001$; see Figures S1-S3 for NC characterization. To collect *in situ* scattering patterns, we flow the emulsion in a closed loop through a quartz capillary aligned with the X-ray beam; see Figure S4 for a schematic of the setup.[28,57] As the volatile oily phase evaporates, $\phi$ increases with time. Figure 1a and Movie S1 illustrate the continuous kinetic evolution of the structure factor, $S(q)$. Initially, the structure factor is featureless and centered around 1, as expected for a colloidal gas. During the first 3 hours drying the emulsion, broad features arise across the wave vector range, $q$. After 3.6 hours, a succession of sharp peaks suddenly emerges from the background, growing in intensity while shifting with time towards higher $q$. As highlighted in Figure 1b, the shape of the structure factor at 3.6 hours resembles that of a low-density colloidal fluid, but within 0.1 hours rapidly develops into a fully formed diffraction pattern featuring at least 7 sharp peaks from growing crystallites. Immediately thereafter, all peaks shift synchronously toward higher $q$, indicating a contraction of the crystal lattice. We identify the crystal as isostructural to $AlB_2$ with parameters $a = b \approx c$; see Figure 1c. This structure is characterized by stacked hexagonal layers of the larger NCs intercalated by hexagonal layers of the smaller NCs that occupy the trigonal prismatic voids left by the larger NCs; see schematic in Figure S5.

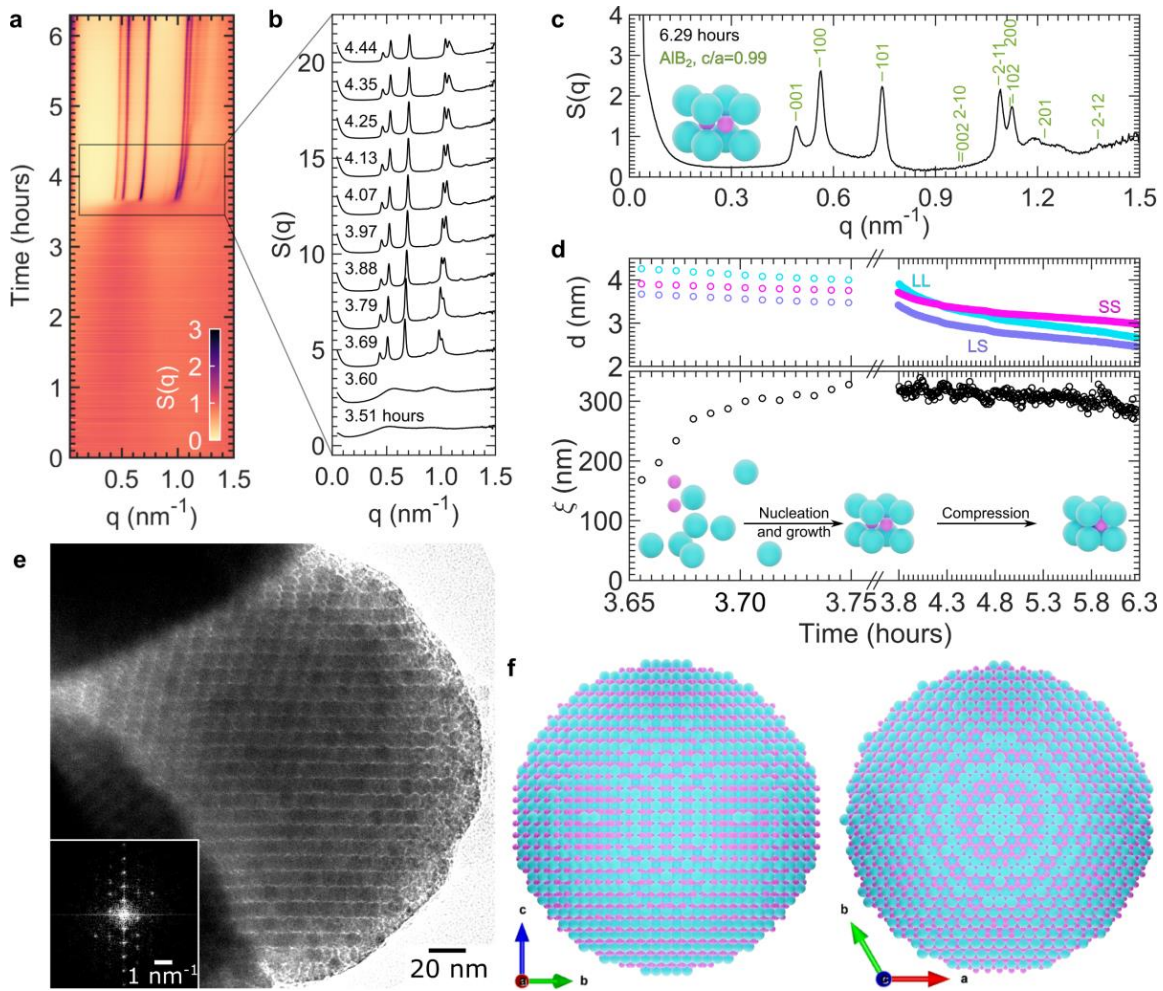

**Figure 1: Formation of colloidal AlB$_2$ BNSLs.** (**a**) Kinetic structure factor, S(q), of a binary dispersion of PbS and Fe$_3$O$_4$ NCs under spherical confinement of a drying emulsion. (**b**) S(q) patterns showing the emergence and evolution of diffraction peaks around the time of BNSL nucleation. (**c**) Final S(q) pattern identifying the BNSL structure as AlB$_2$. (**d**) Evolution of the crystalline lattice after nucleation, highlighting the kinetics of surface-to-surface distance between larger (L) and smaller (S) NCs, $d$ (top), and average crystal size, $\xi$ (bottom). The proposed assembly mechanism is shown as inset. (**e**) TEM micrograph of PbS and Fe$_3$O$_4$ NCs crystallized into a 3D AlB$_2$ BNSL. The fast-Fourier transform is shown as inset. (**f**) Model of the AlB$_2$ BNSL shown in (e).

To understand the formation of these BNSLs, we extract the evolution of the structural parameters from *in situ* measurements as shown in Figure 1d. After nucleation, lattice contraction induces a slow decrease in the surface-to-surface distance between NCs, $d$. This decrease takes place over several hours to reach an inorganic volume fraction of $\phi = 0.357$; see Figure S6. In stark contrast to this steady compression of the lattice, the average crystal size extracted from the Scherrer equation,[62] $\xi$, increases rapidly: Within 0.1 hours after nucleation, the crystal size increases to reach $\xi \approx 330$ nm, corresponding to $\xi/a \sim 23$ unit cells of the BNSL. Eventually, the crystal size slowly decreases to $\xi \approx 280$ nm as the result of lattice compression. Interestingly, we notice that the integrated intensity of the 001 and 100 reflections grow at different rates relative to the 101 reflection, see Figures 1b-c. While the relative integrated intensity of the 001 reflection increases by 63% during assembly, that of the 100 reflection increases by 140%, as shown in Figure S7. This suggests a strong tendency of the AlB$_2$ crystal to grow along the basal plane rather than along the $c$ axis.

Based on these observations, we hypothesize the synthetic mechanism shown in the inset of Figure 1d: Crystallization occurs as a single-step transition from the fluid to the crystalline phase. The relative positions of the diffraction peaks do not change during assembly, implying the absence of intermediate phases between the fluid and the final crystal. This is a simpler process compared to previous reports on the synthesis of single-component NC superlattices reporting crystal-to-crystal transitions.[30,31,52] Crystallization is followed by the continuous compression of the BNSL. We attribute this compression to the evaporation-driven desorption of solvent from the ligand shell of the NCs, consistently with reports on single-component systems.[28,30,57] *Ex situ* TEM confirms the synthesis of 3D colloidal BNSLs, each featuring a distinct set of NC planes, as shown in Figures 1e, S8-S10. The Fourier transform of the BNSL shown in Figure 1e reveals a discrete set of spots, as expected for a crystalline structure. Combining *in situ* and *ex situ* experimental results leads to the structural model shown in Figure 1f. The synthesized BNSLs retain the superparamagnetic character of their larger NC component, while also displaying a reduced magnetic remanence and spin density relative to the single-component superlattice, as shown in Figure S11.

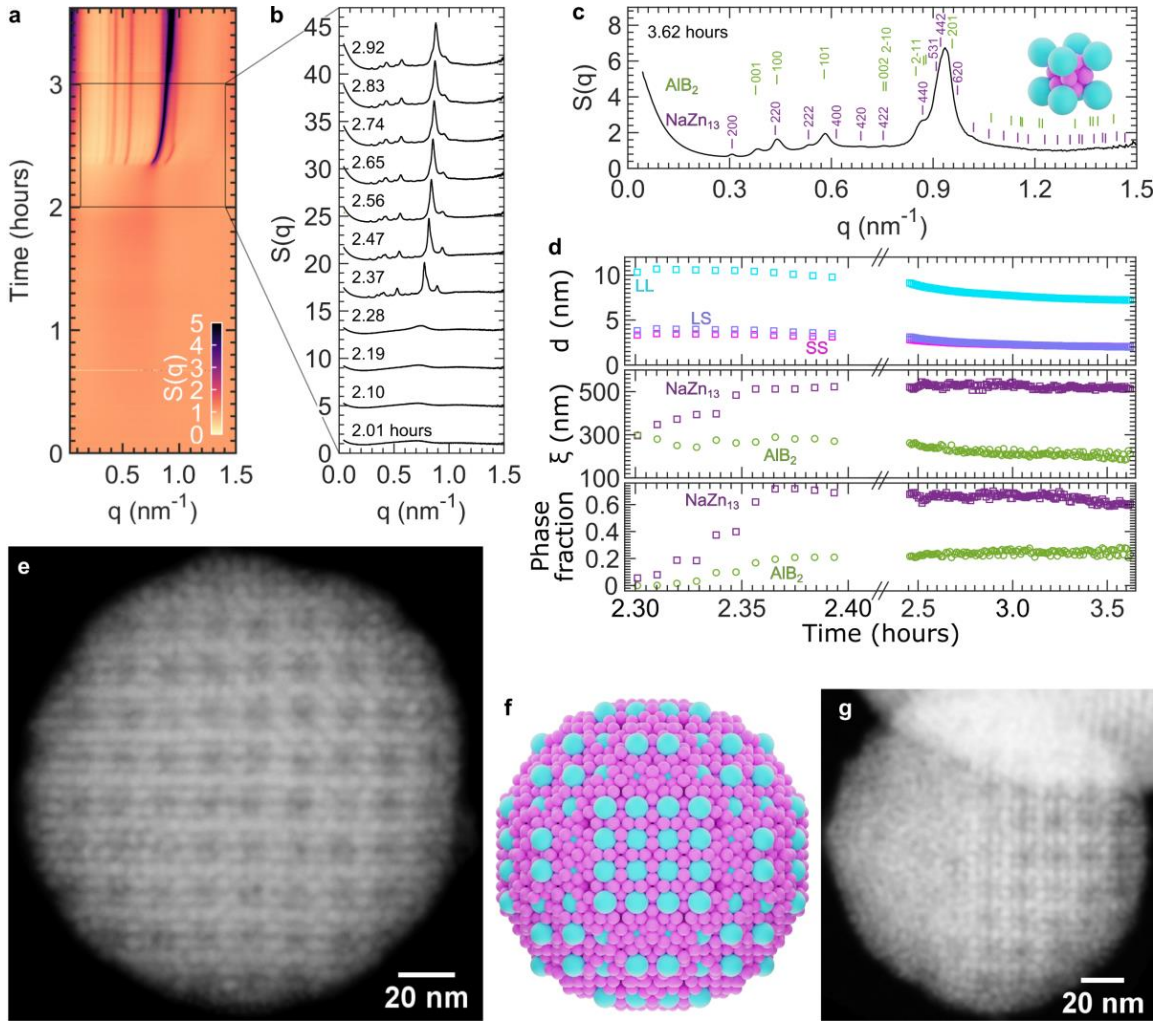

**Figure 2: Formation of colloidal NaZn$_{13}$ BNSLs.** **(a)** Kinetic structure factor, S(q), of a binary dispersion of PbS and FICO NCs under spherical confinement of a drying emulsion. **(b)** S(q) patterns showing the emergence and evolution of diffraction peaks around the time of BNSL nucleation. **(c)** Final S(q) pattern identifying the BNSL structure as NaZn$_{13}$ with a minority AlB$_2$ phase. **(d)** Evolution of the surface-to-surface distance between larger (L) and smaller (S) NCs, $d$ (top), average crystal size, $\xi$ (middle), and phase fraction (bottom) of the two binary phases during lattice compression. **(e)** Dark-field STEM micrographs of a single-crystal NaZn$_{13}$ BNSL, and **(f)** its structural model. **(g)** Micrograph of a heterostructure showing the coexistence of NaZn$_{13}$ with a secondary phase.

We test the robustness of this approach by using a NC pair with comparable size ratio, number ratio, and initial volume fraction but featuring larger infrared plasmonic CdO NCs co-doped with fluorine and indium (FICO) and smaller infrared semiconductor PbS NCs (optical activity in Figure S3). Under the same synthetic conditions, their assembly in emulsion yields the same $AlB_2$ BNSL structure with similar kinetics; see Figure S12 and Movie S2. We next target a different BNSL, $NaZn_{13}$, by increasing the number ratio of these FICO and PbS NCs from 1:2 to 1:13. The experimental structure factor reveals the onset of diffraction peaks from the flat background shortly after 2.3 hours of drying, as shown in Figure 2a-b and Movie S3. The diffraction pattern appears qualitatively different from Figure 1, with at least 10 discernible reflections. A more careful examination of the final diffraction pattern indicates the coexistence of a majority phase isostructural to $NaZn_{13}$ with a minority $AlB_2$ phase, as shown in Figure 2c. The $NaZn_{13}$ structure consists of a body-centered icosahedral cluster of 13 smaller particles contained within a simple cubic subcell of the larger particles, as illustrated in the inset in Figures 2c and S13.

We study the synchronous evolution of the $NaZn_{13}$ and $AlB_2$ phases in Figure 2d. The lattice parameters of the two crystalline phases slowly decrease as a function of time, to reach maximum inorganic volume fractions of *ϕ = 0.360* and *0.384* for $NaZn_{13}$ and $AlB_2$, respectively. Within 0.1 hours after nucleation, the average domain sizes of the $NaZn_{13}$ and $AlB_2$ phases rapidly increase and saturate at their final values of *ξ ≈ 510 nm* and *200 nm*, respectively. We quantify the fraction of each crystalline phase by comparing with the assembly performed at a NC number ratio of 1:2, shown in Figure S12. After nucleation, the fraction of both $NaZn_{13}$ and $AlB_2$ phases quickly increases to reach the values of 0.74 and 0.21, respectively, confirming $NaZn_{13}$ as the majority phase. Interestingly, while the fraction of the $AlB_2$ phase shows a slow increase in the late stages of the assembly, that of the $NaZn_{13}$ phase shows a comparable decrease. This suggests that even though the NaZn$_{13}$ phase readily nucleates to occupy most of the available volume, this structure might be thermodynamically less stable than AlB$_2$. The relative strengths of inter-NC interactions may be responsible for shifting this equilibrium towards one specific phase.

*Ex situ* dark-field STEM confirms the formation of 3D colloidal NaZn$_{13}$ BNSLs; see Figure 2e. The [200] projection clearly illustrates the cubic symmetry of the NaZn$_{13}$ phase, as well as the four-fold symmetry of the smaller NCs surrounding each larger NC; see also Figures S13-S15. In Figure 2f we illustrate a 3D model of this colloidal BNSL, obtained by carving a sphere out of a NaZn$_{13}$ lattice with experimentally determined structural parameters. We also observe superstructures characterized by a heterostructure of NaZn$_{13}$ with a second phase, shown in Figure 2g, confirming the presence of a minority phase as suggested by SAXS. Due to interference with the adjacent NaZn$_{13}$ phase, we are unable to clearly identify this secondary phase using STEM. However, the results of *in situ* SAXS suggest the secondary phase should correspond to AlB$_2$.

*In situ* measurements draw a detailed picture of the self-assembly process that we juxtapose with simulations to reveal the driving force behind the synthesis of BNSLs. The formation of AlB$_2$ and NaZn$_{13}$ BNSLs is frequently attributed to entropy[63,64] because of their high packing fractions[65,66] and because these are both equilibrium phases of the hard-sphere model,[67] whose phase behavior is solely dictated by entropy. However, we recently discovered that the self-assembly of the AlB$_2$ phase is kinetically prohibited in hard-sphere mixtures at a NC number ratio of 1:2,[68] indicating that more complex interactions are necessary for its formation.

The structure of the resulting BNSLs is largely determined by the effective interactions between NCs in solution. The NCs used in this work interact through solvent-mediated van der Waals interactions and superparamagnetic interactions in the case of Fe$_3$O$_4$ NCs.[69] Although many studies assume purely repulsive interactions in rationalizing observed phases in BNSLs (as mentioned above), the fact that our experiments find $AlB_2$, yet $AlB_2$ does not form from purely repulsive NCs at the number ratio 1:2 used in the experiments, implies attraction cannot be neglected.[68] Examples of NC interaction potentials with a repulsive core and attractive well have been obtained from of NC interaction potentials with a repulsive core and attractive well have been obtained from atomistic simulations and theoretical models of similar spheroidal NCs, but the specific shape of the interaction potential varies among systems.[70-75] We choose to model NC interactions with a Mie potential,[76] which has a repulsive core and an attractive well whose depth and width control the strength and range of attraction, respectively. To parameterize the potential, we consider the Noro-Frenkel law of corresponding states,[77] which dictates that only attractive wells with widths above a certain size will have stable liquid phases in their phase diagram. To compare cases with and without stable liquid phases, we examine two different well widths, described as "narrow well" and "wide well", with widths less than and greater than that size, respectively. We show the narrow and wide well in Figure 3a-b, scaling the range of the interaction between two NCs by the average of their effective sizes to account for size differences. For simplicity, the well depth, $\varepsilon$, is set equal for all interspecies interactions, *i.e.* $\varepsilon_{LL} = \varepsilon_{LS} = \varepsilon_{SS} = \varepsilon$. While this is an approximation, large differences in well depths frequently result in demixing,[78] which has been reported for other experiments on binary nanocrystal self-assembly but not those conducted here. Later, we discuss the plausible case of $\varepsilon_{LL} > \varepsilon_{LS} > \varepsilon_{SS}$, finding similar results to the case of equal attraction strength. We also examined a case in which the attractive well between the smaller NCs was significantly narrower than between large NCs, again finding a similar result; see Figure S16.

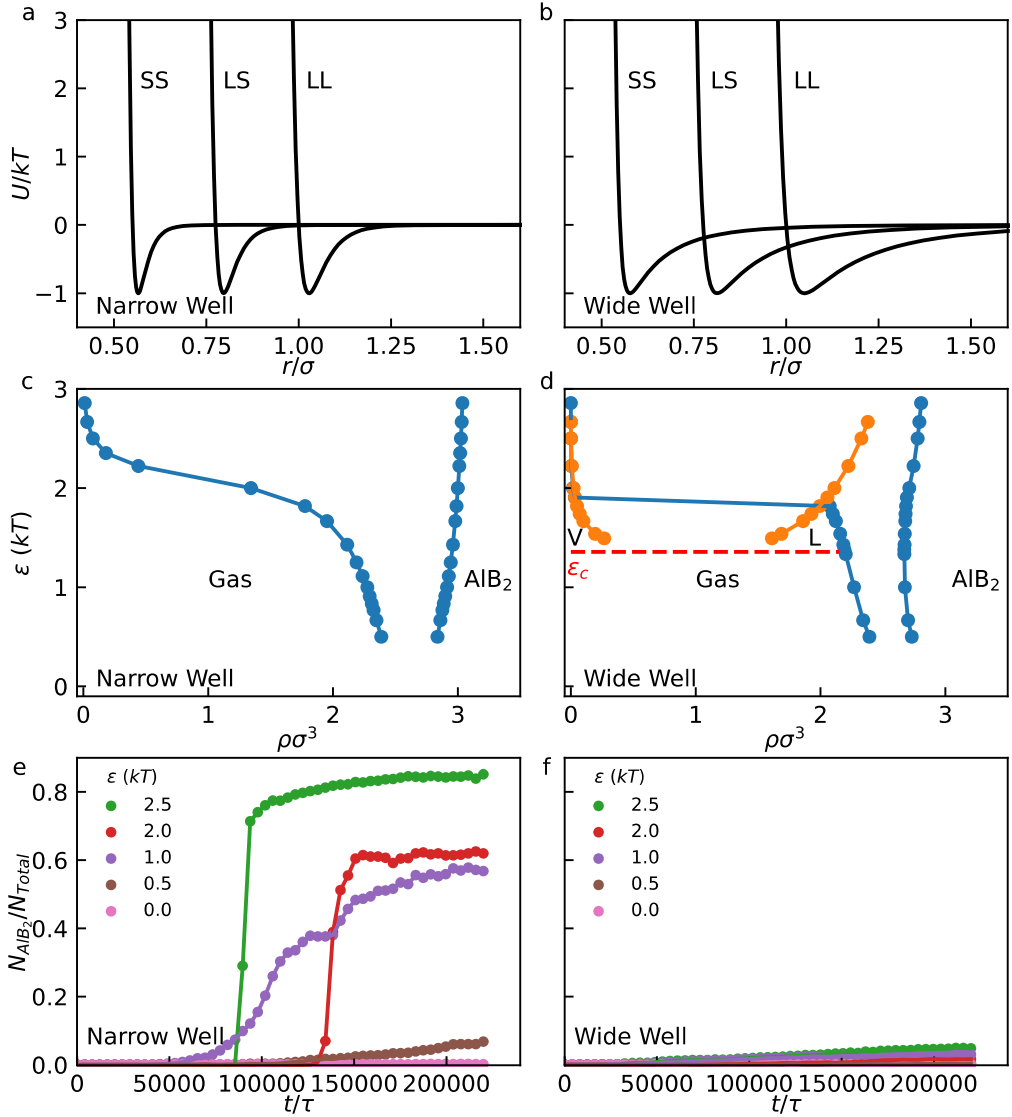

**Figure 3: The influence of attractive forces in binary mixtures. (a-b)** Mie pair potentials shown as a function of the normalized interparticle distance, $U(r/\sigma)$, calculated for a well depth of 1 kT and for length scale parameters $m = 25$ (a) and $m = 6$ (b). The potential in (a) is described as the "narrow well" and the one in (b) as the "wide well" in the main text. For each potential, three types of interactions are shown: between larger NCs (LL), between larger and smaller NCs (LS), and between two smaller NCs (SS). **(c-d)** Thermodynamic phase diagrams computed from free energy calculations for the narrow (c) and wide well (d) as a function of well depth, $\varepsilon$, and normalized particle density, $\rho\sigma^3$. The blue and orange lines demarcate the regions of gas-solid coexistence and vapor-liquid coexistence, respectively. Errors in the phase boundaries are smaller than the points. The dashed red line in (d) indicates the critical well depth $\varepsilon_c$ above which vapor-liquid coexistence occurs. The phase diagrams are computed at a NC number ratio of 1:2. **(e-f)** The evolution of the number of AlB$_2$-like particles in self-assembly simulations through slow compression for the narrow (e) and wide well (f).

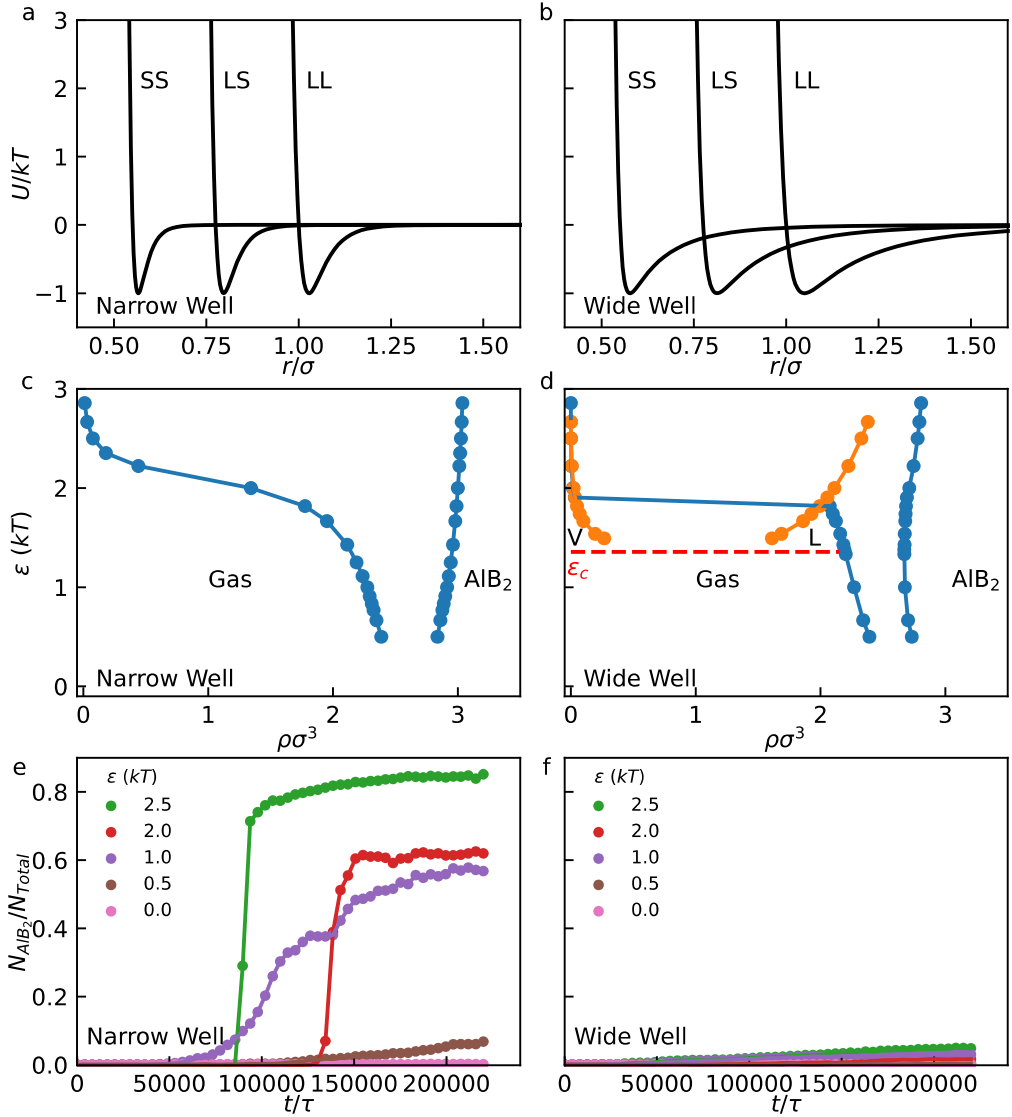

**Figure 3: The influence of attractive forces in binary mixtures. (a-b)** Mie pair potentials shown as a function of the normalized interparticle distance, $U(r/\sigma)$, calculated for a well depth of 1 kT and for length scale parameters $m = 25$ (a) and $m = 6$ (b). The potential in (a) is described as the "narrow well" and the one in (b) as the "wide well" in the main text. For each potential, three types of interactions are shown: between larger NCs (LL), between larger and smaller NCs (LS), and between two smaller NCs (SS). **(c-d)** Thermodynamic phase diagrams computed from free energy calculations for the narrow (c) and wide well (d) as a function of well depth, $\varepsilon$, and normalized particle density, $\rho\sigma^3$. The blue and orange lines demarcate the regions of gas-solid coexistence and vapor-liquid coexistence, respectively. Errors in the phase boundaries are smaller than the points. The dashed red line in (d) indicates the critical well depth $\varepsilon_c$ above which vapor-liquid coexistence occurs. The phase diagrams are computed at a NC number ratio of 1:2. **(e-f)** The evolution of the number of AlB$_2$-like particles in self-assembly simulations through slow compression for the narrow (e) and wide well (f).

We note that our model assumes that more complex interactions, like many-body or anisotropic effects, either do not play a large role in determining the phase behavior of these NCs or they conspire to produce an effective interaction similar to the model. If we detect in simulation the phases observed in experiment while neglecting these more complex interactions, we can then infer that the simpler interactions are likely responsible for the formation of these phases. Furthermore, more complex interactions such as dipolar interactions[79,80] are more likely to be influenced by the specifics of the NC and solvent compositions, but the AlB$_2$ and NaZn$_{13}$ phases have been reported for many NC and solvent compositions,[21,22,63,78-82] reducing the likelihood that composition-dependent interactions are necessary for their self-assembly.

To reveal how the range and strength of interaction affect thermodynamic phase behavior, we compute phase diagrams for a NC number ratio of 1:2, as shown in Figure 3c-d. In the presence of the narrow well, the gas and solid $AlB_2$ phases are both stable. In contrast, in the presence of the wide well a region of vapor-liquid coexistence is stable above a critical well-depth $\varepsilon_c$; see Figures 3d and S17. This is consistent with the aforementioned Noro-Frenkel law of corresponding states,[77] which predicts that wider potential wells exhibit a stable liquid–gas transition.

While these phase diagrams show the equilibrium predictions for a given set of parameters, they do not indicate whether a phase is kinetically accessible. To study whether the $AlB_2$ phase will form, we slowly compress an initially disordered fluid under periodic boundary conditions. A combination of Steinhardt order parameters enables the quantification of the fraction of larger NCs that become crystalline as a function of time, $N_{AlB_2}/N_{Total}$; see Figure S18 for calculation details.[83] When using a narrow well at least 1.0 kT deep, over 50% of larger NCs crystallize by the end of the simulations, as shown in Figure 3e. There is limited crystallization for a shallower well of 0.5 kT, and no crystallization when the NCs are purely repulsive. In contrast, only minimal crystallization occurs with the wide well, with at most 6% of larger NCs registering as crystalline even for the deepest well investigated, 2.5 kT, as shown in Figure 3f.

These results show that a deep, narrow attractive well substantially improves the crystallization kinetics of $AlB_2$. A similar kinetic enhancement has been reported in simulations of single-component systems (as discussed below), but to our knowledge this is the first demonstration of such an effect in a multicomponent system of this nature. Furthermore, the effect is more dramatic for the case of AlB$_2$, because AlB$_2$ fails to crystallize at all without an attractive well but crystallizes readily with one. In comparable single-component systems, a narrow attractive well can improve crystallization kinetics, but self-assembly still occurs readily without one (i.e. when the NCs resemble hard spheres).

We can rationalize this improvement to kinetics by considering the phase diagrams shown in Figures 3c-d. In general, crystal nucleation and growth is strongly influenced by the degree of supersaturation and particle mobility. In the phase diagram shown in Figure 3c, the solid become stable at low densities (where the particles are highly mobile) for a narrow, deep attractive well. A similar stabilization of the solid phase occurs for the wide well system, but the dense liquid phase is also stabilized, reducing the supersaturation. We quantify these effects by computing the chemical potential driving force between the crystal and fluid phases as a function of NC mobility in Figure S19, finding that deep, narrow wells do result in higher chemical potential driving forces at higher particle diffusivities. We also note that this principle of achieving high driving forces at low density is similar to that proposed for colloids with small attractive patches.[84]

To place these observations in a more general context, we investigated the crystallization of two related systems: a single-component face-centered cubic crystal (FCC) and the two-component NaZn$_{13}$ crystal for the narrow well at different well depths. In Figure S20, we show that, as with AlB$_2$, crystallization of both structures occurs at increasingly lower densities, although for FCC this shift is not too important because self-assembly still occurs readily without an attractive well. Furthermore, above its critical well depth, we find that AlB$_2$ crystallized through a two-step process in which a dense liquid forms first; see Figures S21 and S22. This switch to a two-step nucleation process at the critical point is consistent with previous reports for single-component systems. Thus, we conclude that crystallization in our binary systems occurs very similarly to that in single-component systems, except that an attractive force is required to observe the formation of some binary crystals.

In summary, our results indicate that the interaction between NCs during self-assembly is consistent with a pair potential characterized by a narrow well. We note that such a pair potential is fairly common for colloids in which vdW forces are the largest contribution to inter-NC attraction. Furthermore, we note that the narrow attractive well we use closely resembles that used for similar NCs in a recent study in which the attraction between NCs was assumed to be entirely vdW interactions between NC cores.[29] We also note that $AlB_2$ will self-assemble for a broad range of well depths greater than 0.5 kT, which possibly explains why it self-assembles for a NCs with different compositions, including the CdO NCs used here.

We next checked for further consistency with experiment by simulating conditions closer to experiment by adding NC polydispersity and spherical confinement to the simulations. We determine the polydispersity of each NC species by SAXS, Figure S1, and simulate the interaction of each NC with the confining boundary of the droplet using a Weeks-Chandler-Anderson potential.[85] Unsurprisingly, we found that adding polydispersity disfavored crystallization, but that crystallization was still possible on our time scales for the highest well depth examined in Figure 3 (2.5 kT), so we used that well depth in these simulations. Each simulation system is initialized as a colloidal fluid in a droplet, then slowly compressed to induce self-assembly. In close agreement with experiment, $AlB_2$ forms at a NC number ratio of 1:2, while $NaZn_{13}$ forms at 1:13. Interestingly, crystallization begins at a slightly lower volume fraction for $NaZn_{13}$ than $AlB_2$, 0.528 and 0.572 respectively, as shown in Figure 4a. Simulations performed at lower volume fractions did not result in self-assembly. As shown by the kinetic change in the fraction of crystalline particles, rapid crystal growth follows nucleation. Consistent with experiments, we find no intermediate phases that precede the final phases, indicating classical nucleation from a dense fluid phase.[30,31] Furthermore, we conclude that spherical confinement does not influence the identity of the self-assembled structure in the systems investigated here.

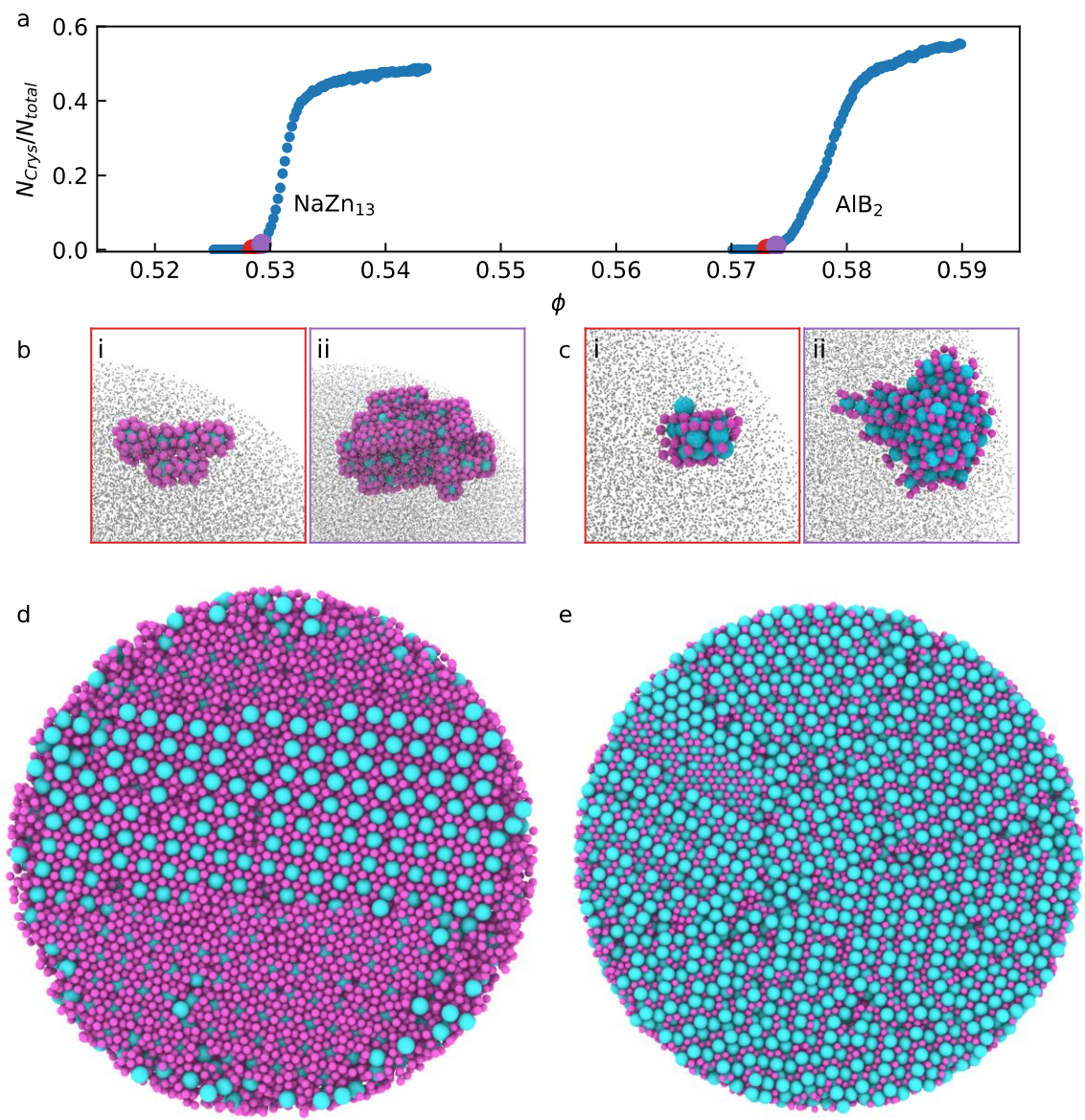

**Figure 4: The self-assembly of AlB$_2$ and NaZn$_{13}$ in spherical droplets.** In (**a**) we show the results of attempts to self-assemble AlB$_2$ and NaZn$_{13}$ with the deepest (2.5 kT), narrow well used in Figure 3. The curve labelled "NaZn$_{13}$" was obtained at a stoichiometry of 1:13, while the curve labelled "AlB$_2$" was obtained at a stoichiometry of 1:2; $N_{crys}/N_{total}$ is the fraction of large particles we identify as NaZn$_{13}$-like or AlB$_2$-like respectively. In (**b**) and (**c**) we show early stages of the growing NaZn$_{13}$ and AlB$_2$ crystals, respectively. We show two different time points, (**i**) and (**ii**); the color of the images' borders match that of the corresponding timepoints in (a) from which they were taken. We show large particles identified as crystalline in blue and small particles neighboring a crystalline large particle in pink; all other particles are reduced in size and colored grey. In (**d**) and (**e**), we visualize an inner slice of each droplet at the final time point of our simulations, coloring every particle. Unlike the simulations in Figure 3, each species is set to have the polydispersity of the corresponding experimental nanoparticles.

Simulations provide the unparalleled advantage of directly visualizing crystal nucleation,[86-88] a process notoriously elusive to capture in experiments. The early stages of assembly for $NaZn_{13}$ and $AlB_2$ BNSLs are shown in Figures 4b and 4c, respectively. To highlight the crystalline nuclei, we color only the NCs identified as being in a crystalline environment, while fluid-like NCs are shown as smaller grey spheres. For both $NaZn_{13}$ and $AlB_2$ BNSLs, the critical nuclei emerge from the fluid multiple particle diameters away from the surface of the droplet, allowing us to conclude that these BNSLs undergo homogeneous nucleation. Only after nucleation do the crystallites of $NaZn_{13}$ and $AlB_2$ BNSL spread to the wall. This behavior is unaffected by the size of the droplet, as shown by replacing the spherical walls with flat walls to simulate significantly larger droplets, as shown in Figure S23. We find no evidence for exotic pre-nucleation clusters.[63,89]

The final stages of growth result in the crystals shown in Figures 4d-e. The $NaZn_{13}$ grains are easily identifiable by the simple cubic arrangement of the larger NCs. A single crystal grain spans most of the spherical superstructure, consistent with the experimental results shown in Figure 2. In contrast, multiple grains of $AlB_2$ are present. This qualitatively agrees with the SAXS measurements showing smaller grains for $AlB_2$ than $NaZn_{13}$, although specific crystal grains are harder to visualize in the TEM micrographs, making a quantitative comparison between simulation and experiment challenging.

An apparent discrepancy remains between experiment and simulation: experimental results indicate the presence of ~20% $AlB_2$ as a second phase for samples prepared at a NC number ratio of 1:13, while simulations show less than 1% $AlB_2$. Interestingly, reducing the magnitude of the attraction between smaller NCs results in the coexistence of $AlB_2$ and $NaZn_{13}$ at a NC number ratio of 1:13, removing the discrepancy as shown in Figure S24. This adjustment is consistent with the dependence of van der Waals and superparamagnetic interactions on NC size.[90]

**Conclusions.**

We show a remarkable correspondence between experiment and simulation on the synthesis of BNSLs through self-assembly. While early efforts using evaporation-driven self-assembly have revealed the tendency to nucleate multiple polymorphs simultaneously,[22,79,91-93] our results show that emulsion-templated assembly provides a more controlled pathway towards the generation of phase-pure BNSLs. Under spherical confinement, NCs readily nucleate into binary phases isostructural to $AlB_2$ and $NaZn_{13}$ without intermediate liquid or crystal phases. The burst of crystal nucleation is followed by a gradual lattice contraction to result in multifunctional, 3D, dense, crystalline binary phases. We can accurately reproduce these experimental results in simulation by introducing a short-ranged, attractive potential which kinetically promotes self-assembly. This direct link between experiments and simulations reveals that BNSLs nucleate homogeneously and directly, without intermediate solid phases preceding the final crystal. Aspects of our results likely apply to NCs coated with DNA-based ligands,[40,94-96] which may provide more continuous control over the range of interactions and thus allow for a more direct probing of the remaining open questions. In achieving a closer correspondence between experiment and simulation, and demonstrating the importance of short-range attraction for assembly kinetics, this work represents a crucial first step in *a priori* prediction of BNSLs towards the deterministic hetero-integration of NCs into multifunctional structures, targeting applications in photonics,[53,58-60,97] excitonics,[58,98] phononics,[99,100] and catalysis.[101,102]

**Methodology**

*Experiments*

  **Synthesis and characterization of the NC building blocks**: Oleate-functionalized PbS (lead sulfide, rock-salt structure, 4.5 ± 0.4 nm and 6.4 ± 0.6 nm in diameter), $Fe_3O_4$ (iron oxide, cubic spinel structure, 10.4 ± 0.6 nm in diameter), and FICO (fluorine and indium co-doped cadmium oxide, rock-salt structure, 13.2 ± 0.9 nm in diameter) NCs were synthesized by following reported procedures and redispersed in toluene.[101,103,104] The NC concentration was determined either by spectrophotometry by using a published sizing curve[105] (PbS, Figure S3) or by weighing the dry pellet ($Fe_3O_4$ and FICO). The size of the inorganic cores of the NCs was determined by fitting the SAXS pattern measured from a dilute dispersion of NCs with a spherical form factor, Figure S1. The form factor was convoluted with a Gaussian distribution of sizes to account for NC polydispersity. The fitting was performed by using the free software SASfit.[106] The effective size of the NCs was determined by imaging a monolayer of NCs by TEM, calculating the fast-Fourier transform of the image, and extracting the center-to-center distance between nearest neighbors, Figure S2. A complete description of these procedures is available in the supplementary information.

  *In situ* **SAXS**: The kinetic patterns were collected at the SMI beamline, Brookhaven National Laboratory, using a recently developed experimental setup.[28] To conserve beamtime, we expanded the setup to support the simultaneous measurement of 4 samples by translating the sample stage vertically; see schematic in Figure S4. Each sample was prepared as follows: A 20 mL scintillation vial was charged with 8 mL of 200 mM sodium dodecyl sulfate in water. Subsequently, the vial was charged with 2 mL of a NC dispersion in 22 v/v % toluene and 78 v/v % hexanes with a total NC volume fraction of 0.001. The vial was capped and vigorously vortexed for 60 seconds using a vortex mixer (Fisher) to generate the emulsion. The emulsion was then uncapped and diluted by adding 10 mL of 200 mM sodium dodecyl sulfate in water. A 1-inch octagonal stir bar was then added to the diluted emulsion. The vial was placed on a hotplate (IKA plate) equipped with a thermocouple and a heating block for vials, heated to 70 °C while stirring at 500 rpm, and allowed to flow by means of a peristaltic pump (Cole-Palmer) at a flow rate of 10 mL/minute through a closed loop of Viton peristaltic tubing (Cole-Palmer). The closed loop included a custom-made flow cell consisting of a 1 mm quartz capillary tube (Charles Supper). The X-ray beam was aligned with the center of the capillary. This setup allowed us to measure the scattering pattern from the emulsion as evaporation occurred from the uncapped vial. The integration time for each measurement was set to 1 second, the time between consecutive acquisitions was 28 seconds, the beam energy was 16.1 keV, and the sample to detector distance was 6.3 meters. The q-range was calibrated against a silver behenate standard. The two-dimensional patterns were azimuthally averaged, and background subtracted to yield $I(q,t)$, where $t$ is the time. The kinetic structure factor, $S(q,t)$, was then obtained by calculating $S(q,t) = I(q,t)/I(q,0)$ since at the beginning of the experiment the NCs are well dispersed within the droplets.

**Structural parameters of AlB$_2$**: For a hexagonal structure (Figure S5), the expected reflections $q_{hkl}$ for the planes of indexes $hkl$ are:

$$q_{hkl} = \sqrt{\frac{4}{3}\frac{h^2 + hk + k^2}{a^2} + \frac{l^2}{c^2}}$$

We first extracted the $c/a$ factor for the AlB$_2$ structure for the last data point, $t = t_{end}$. We did so empirically by generating the expected positions $q_{hkl}$ of the reflections for a given value of $c/a$, and comparing with the experimental results until an agreement was found. We found that the value of $c/a$ does not vary as the structure evolves during drying. To calculate the kinetic structural parameters, we first fitted each $S(q,t)$ curve with a superposition of Lorentzian curves with line shape:

$$L(q) = \frac{Aw^2}{(q-q_{hkl})^2 + w^2}$$

centered around the expected $q_{hkl}$ positions, with amplitude $A$ and full width at half maximum $2w$. Since $q_{101}$ was the most isolated reflection, we used it to calculate the kinetic structural parameters. From the Scherrer equation, we calculated the average crystal size as $\xi = 2\pi K/2w = \pi K/w$, where $K = 1.0747$ is the Scherrer constant used for a spherical crystal.[107] The lattice parameter, also equal to the center-to-center distance, or bond length, between larger (L) NCs was calculated as:

$$b_{LL} = a = \frac{2\pi}{q_{101}}\sqrt{\frac{4}{3} + \frac{1}{(c/a)^2}}$$

The surface-to-surface distance between larger NCs was then calculated as $d_{LL} = b_{LL} - \sigma_L$, where $\sigma_L$ is the average diameter of the inorganic cores of the larger NCs as measured by *ex situ* SAXS. The bond length between smaller (S) NCs was calculated by scaling by the expected values for the bond lengths in the atomic AlB$_2$ structure for which $b_{LL,At} = 0.30090 \, nm$ and $b_{SS,At} = 0.17372 \, nm$ so that $b_{SS} = b_{LL}(b_{SS,At}/b_{LL,At})$. The surface-to-surface distance between smaller NCs was then calculated as $d_{SS} = b_{SS} - \sigma_S$, where $\sigma_S$ is the average diameter of the inorganic cores of the smaller NCs as measured by *ex situ* SAXS. The bond length between larger and smaller NCs was then calculated as $b_{LS} = b_{LL}(b_{LS,At}/b_{LL,At})$, where $b_{LS,At}$ is the bond length in the atomic AlB$_2$ structure for measured value of $c/a$ in the nanocrystal superstructure. For $c/a = 0.99$, $b_{AB,At} = 0.22883$ $nm$. The surface-to-surface distance between larger and smaller NCs was then determined as $d_{LS} = b_{LS} - \sigma_L/2 - \sigma_S/2$. Additional details are provided in the supporting information.

**Structural parameters of NaZn$_{13}$**: For a cubic structure (Figure S13), the expected reflections $q_{hkl}$ for the planes of indexes $hkl$ are:

$$q_{hkl} = \frac{2\pi}{a}\sqrt{h^2 + k^2 + l^2}$$

To calculate the kinetic structural parameters, we first fitted each $S(q,t)$ curve with a superposition of Lorentzian curves with line shape centered around the expected $q_{hkl}$ positions, with amplitude $A$ and full width at half maximum $2w$. Since $q_{200}$ was the most isolated reflection, we used it to calculate the kinetic structural parameters. From the Scherrer equation, we calculated the average crystal size as $\xi = \pi K/w$. The lattice parameter $a$, twice the center-to-center distance, or bond length, between larger NCs, $b_{LL}$, was calculated as $2b_{LL} = a = 4\pi/q_{200}$. The surface-to-surface distance between larger NCs was then calculated as $d_{LL} = b_{LL} - \sigma_L$. The bond length between smaller NCs was calculated by scaling by the expected values for the bond lengths in the atomic NaZn$_{13}$ structure for which $b_{LL,At} = 0.61365$ $nm$ and $b_{SS,At} = 0.25664$ $nm$ so that $b_{SS} = b_{LL}(b_{SS,At}/b_{LL,At})$. The surface-to-surface distance between smaller NCs was then calculated as $d_{SS} = b_{SS} - \sigma_S$. The bond length between larger and smaller NCs was then calculated as $b_{LS} = b_{LL}(b_{LS,At}/b_{LL,At})$ where $b_{LS,At} = 0.35647$ $nm$ is the bond length in the atomic NaZn$_{13}$ structure.

The surface-to-surface distance between larger and smaller NCs was then determined as $d_{LS} = b_{LS} - \sigma_L/2 - \sigma_S/2$. Additional details are provided in the supporting information.

**Electron microscopy**: For TEM and STEM imaging, a JEOL F200 microscope was operated at 200 kV. During imaging, magnification, focus and tilt angle were varied to yield information about the crystal structure and super structure of the particle systems. To prepare the superstructures for imaging, after the emulsion had fully dried, the binary NC superstructures were washed twice in a solution of 20 mM sodium dodecyl sulfate in water by centrifugation at 3000 g for 30 minutes, and redispersed. 10 μL of the dispersion was drop casted on a carbon-coated TEM grid (EMS) and dried under vacuum for 1 hour. The grid was then dipped in a cleaning solution consisting of 1:2 water:isopropranol by volume, and dried for 1 hour under vacuum.

**Rendering:** The coordinates for a three-dimensional superlattice with parameters matching experiment were initially generated by using a self-developed script. A sphere of a desired size matching a superstructure was then carved from the superlattice. The positions and sizes of all particles were then sent to the free software Blender 2.93 and rendered.

*Simulations*

We used molecular dynamics (MD) with the HOOMD-Blue simulation toolkit[108] to simulate a binary mixture of NCs with interactions modeled by the Mie (IPL) potential:

$$U_{ij}(r_{ij}) = \varepsilon_{ij} \left(\frac{n}{n-m}\right)\left(\frac{n}{m}\right)^{\frac{m}{n-m}}\left(\left(\frac{\sigma_{ij}}{r_{ij}}\right)^n - \left(\frac{\sigma_{ij}}{r_{ij}}\right)^m\right)$$

where $U_{ij}(r_{ij})$ is the energy between two NCs (particles) $i$ and $j$ separated by a distance $r_{ij}$. The potential is described by four parameters: a measure of the particle's size, $\sigma_{ij}$; the power of the repulsive component, $n$; the magnitude of the interaction, $\varepsilon_{ij}$; and the length scale of the attractive interaction, $m$. We set $\sigma_{ij}$ to match the effective size ratio of the NCs used in experiment: $\sigma_{SS} = 0.55 * \sigma_{LL}$, $\sigma_{LS} = \frac{0.55+1}{2} * \sigma_{LL}$, and $\sigma_{LL} = 1$, where $L$ represents the larger NCs and $S$ the smaller NCs. For simplicity, we set the depth of the potential well to be equal for all particle pairs: $\varepsilon_{LL} = \varepsilon_{LS} = \varepsilon_{SS} = \varepsilon$. We also use $\sigma$ to represent $\sigma_{LL}$ in the text. For consistency with previous works,[68] we set the power $n$ to a value of 50. We analyze systems with $m$ of 6 and 25, which we refer to as "wide well" and "narrow well" respectively. The resulting potentials are shown in Figure 3a-b for a well-depth of 1.0 kT. Throughout the paper we manipulate the well depth by changing the temperature, which is inversely proportional to the well depth. We define the units of time as $\tau = \sigma \left(\frac{w}{\varepsilon}\right)^{\frac{1}{2}}$, where $w$ is mass and set to 1 for every type of particle. For $\varepsilon/kT = 0$ shown in Figure 3c-d, we simulated an inverse power law potential at 1.0 kT, similarly to previous work.[68]

To compute the free energies of different phases in Figure 3c and 3d, we combined thermodynamic integration with the Einstein molecule method,[109] a variant of the Frenkel-Ladd method,[110] using at least 2,000 particles in every case. The free energies of the gas and liquid phases were computed at a stoichiometry of 1:2. Self-assembly was attempted with 27,000 particles by slowly compressing the particles from an initially disordered fluid state to a crystalline or kinetically arrested amorphous state. In Figure S25 we give the range of densities compressed over for each well depth and width. We also compute the diffusion coefficient in the vicinity of kinetic arrest. We used NVT simulations based on the MTK equations[111] to thermostat our simulations in Figure 3e-f and NVT simulations using a Langevin integrator[112] to thermostat our simulations in Figure 4.

For the simulations in Figure 4, we treat the distribution of particle sizes as a mixture of two normal distributions: one centered at a size of $1\sigma_{LL}$ and one centered at a size of $0.55\sigma_{LL}$. The standard deviations (*s*) of the normal distributions were chosen to match experiment: $s = 0.047\sigma_{LL}$ for that of the larger particles and $s = 0.063\sigma_{SS}$ for that of the smaller particles. We then discretized the distributions, with 13 bins associated with each peak. We placed the particles inside a spherical droplet, whose edges repel the particles with a Weeks-Chandler-Anderson potential.[85] We computed an effective packing fraction by calculating an effective particle size according to the prescription of Barker and Henderson.[113] We scaled the wall's range of interaction by $\sigma_i/2$, which accounts for the different sizes of the particles.

We used Steinhardt order parameters[83] in Figures 3 and 4 to identify crystalline particles. The specific combinations for each crystal are shown in Figure S18. The parameters were calculated using the *freud* software library.[114] In Figure S22 we used the local density of each particle to infer the occurrence of two-step nucleation. We computed the local density using the implementation provided by *freud*[114], in which the contribution of each neighbor is scaled by the interparticle distance and diameter of the particle. We used $r_{max} = 1.3\sigma$ and a diameter of $1.0\sigma$ in the calculation. We classified particles as locally dense if the local density was greater than $1.05/\sigma^3$, $1.6/\sigma^3$, and $2.5/\sigma^3$ for simulations of FCC, $AlB_2$, and $NaZn_{13}$, respectively. These cutoffs were chosen to separate particles belonging to the initial low-density fluid phase from denser fluid and crystal phases.

We used Ovito[115] to visualize our simulations throughout this work.

The computational workflow and data management for this publication was primarily supported by the *signac* data management framework.[116,117]

AUTHOR INFORMATION

**Corresponding Author**

*email: cbmurray@sas.upenn.edu, sglotzer@umich.edu

**Author Contributions**

E.M. designed the experiment. E.M., S.v.D., A.W.K., and D.A. synthesized and characterized the NC building blocks. E.M., S.v.D., S.Y., D.J.R., and E.H.R.T. measured the *in situ* scattering data. E.M. analyzed the results. E.H.R.T. provided local support at the beamline. E.M., G.G., and S.v.D. performed the electron microscopy studies. D.J.R. performed the magnetic measurements. R.A.L. and T.C.M. performed the simulations and analyzed the results. S.C.G. and C.B.M. supervised the project. The manuscript was written through contributions of all authors. All authors have given approval to the final version of the manuscript. ‡These authors contributed equally.

ACKNOWLEDGMENT

The authors acknowledge primary support from the National Science Foundation under Grant DMR-2019444. E.M. and S.Y. acknowledge support from the Office of Naval Research Multidisciplinary University Research Initiative Award ONR N00014-18-1-2497 for sample preparation and characterization. A.W.K. and C.R.K. acknowledge support from the Semiconductor Research Corporation (SRC) under the Nanomanufacturing Materials and Processes (NMP) trust via Task 2797.001. D.J.R. acknowledges support from the VIEST fellowship. G.G. acknowledges Solvay for financial support. C.B.M. acknowledges the Richard Perry University Professorship at the University of Pennsylvania. Support for the Dual Source and Environmental X-ray Scattering Facility at the University of Pennsylvania was provided by the Laboratory for Research on the Structure of Matter which is funded in part by NSF MRSEC 1720530. This research used resources of the Center for Functional Nanomaterials and the National Synchrotron Light Source II, which are U.S. DOE Office of Science Facilities, at Brookhaven National Laboratory under Contract No. DESC0012704. Computational work used resources form the Extreme Science and Engineering Discovery Environment (XSEDE),[118] which is supported by National Science Foundation grant number ACI-1548562; XSEDE Award DMR 140129. Additional computational resources and services supported by Advanced Research Computing at the University of Michigan, Ann Arbor.

# Supporting Information

for

# *Crystallization of Binary Nanocrystal Superlattices and the Relevance of Short-Range Attraction*

*Emanuele Marino,*[1,8,‡] *R. Allen LaCour,*[4,‡] *Timothy C. Moore,*[4] *Sjoerd W. van Dongen,*[1,6] *Austin W. Keller,*[2] *Di An,*[1] *Shengsong Yang,*[1] *Daniel J. Rosen,*[2] *Guillaume Gouget,*[1] *Esther H.R. Tsai,*[7] *Cherie R. Kagan,*[1,2,3] *Thomas E. Kodger,*[6] *Sharon C. Glotzer,*[4,5,*] *and Christopher B. Murray*[1,2,*]

[1]Department of Chemistry, [2]Department of Materials Science and Engineering, [3]Department of Electrical and Systems Engineering, University of Pennsylvania. Philadelphia, Pennsylvania, 19104, United States. [4]Department of Chemical Engineering and [5]Biointerfaces Institute, University of Michigan, Ann Arbor, Michigan 48109, United States. [6]Physical Chemistry and Soft Matter, Wageningen University and Research, 6708WE, Wageningen, The Netherlands. [7]Brookhaven National Laboratory, Bldg. 735, Center for Functional Nanomaterials, Upton, NY 11973-5000, USA. [8]Dipartimento di Fisica e Chimica, Università degli Studi di Palermo, Via Archirafi 36, 90123 Palermo, Italy.

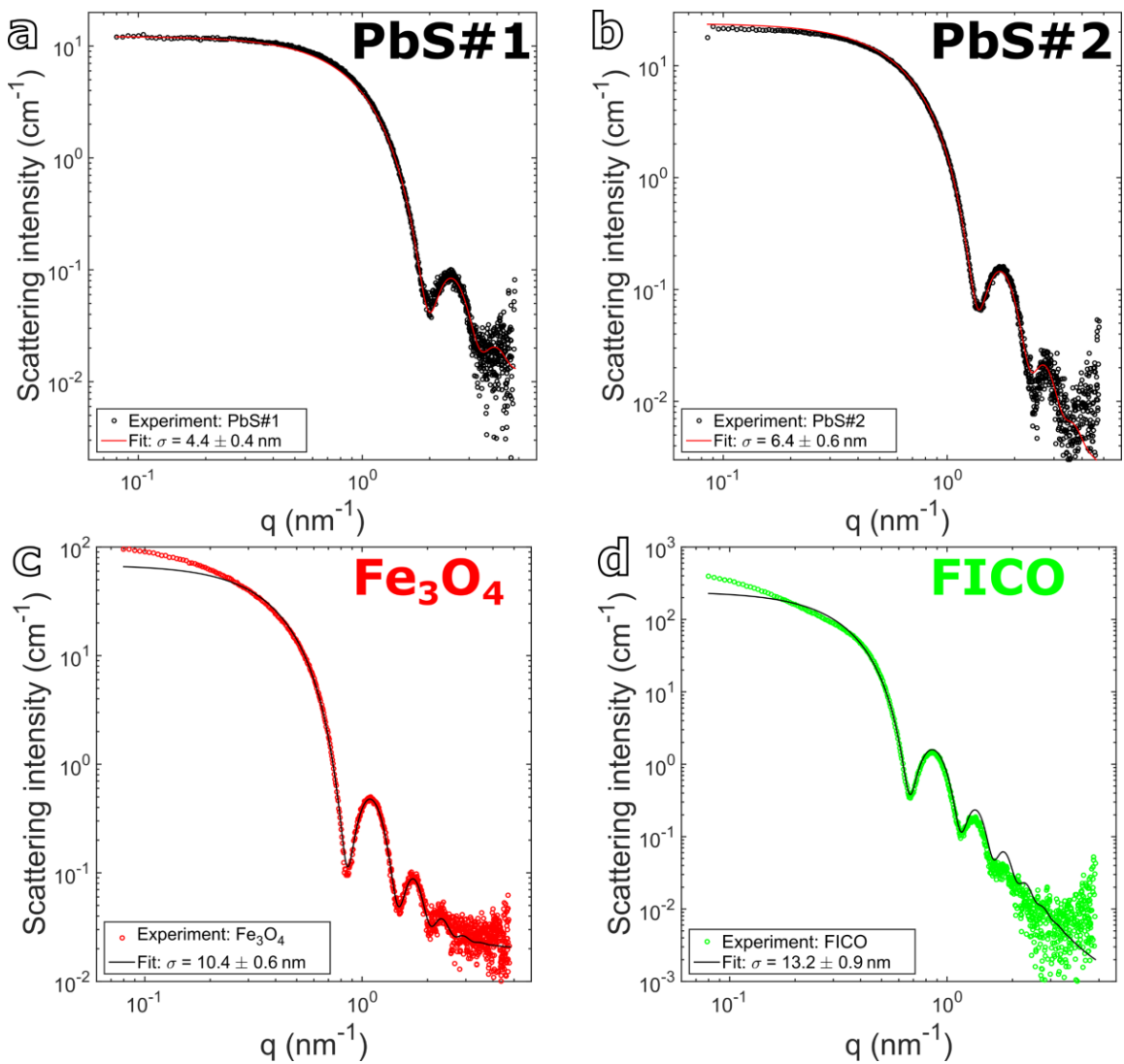

**Figure S1:** Static SAXS patterns of the nanocrystals used in this study. The patterns were collected from 10 mg/mL dispersions in toluene. The intensity was calibrated to absolute units by using the signal from water. The wavevector range was calibrated against a silver behenate standard.

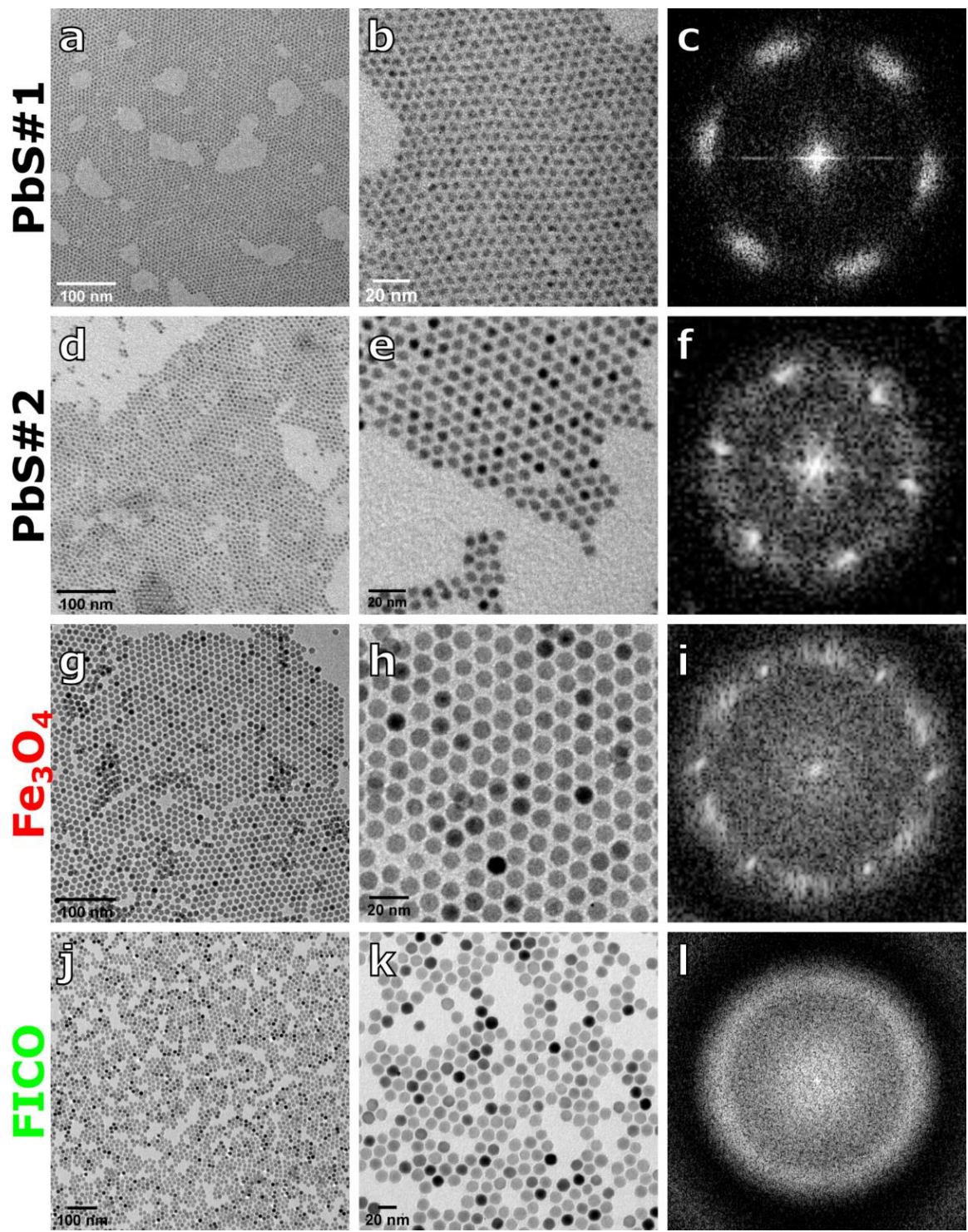

**Figure S2:** TEM micrographs of the nanocrystals used in this study: PbS#1 (a-c), PbS#2 (d-f), $Fe_3O_4$ (g-i), and FICO (j-l). The right-most column represents the FFT image illustrating the method to determine the effective nanocrystal size as described in the supporting text.

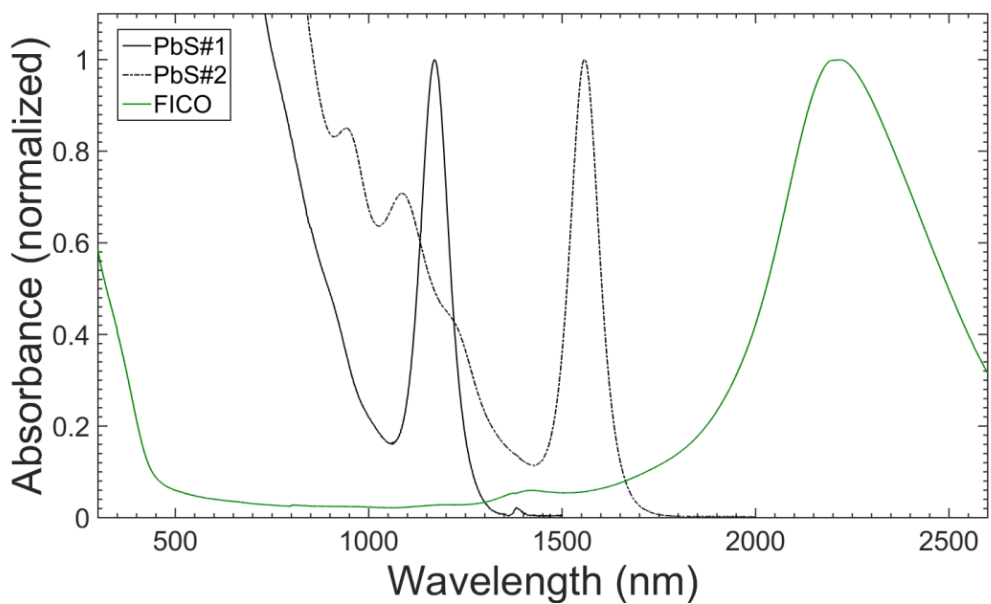

**Figure S3:** Spectrophotometry data from the semiconductor and plasmonic nanocrystals used in this study.

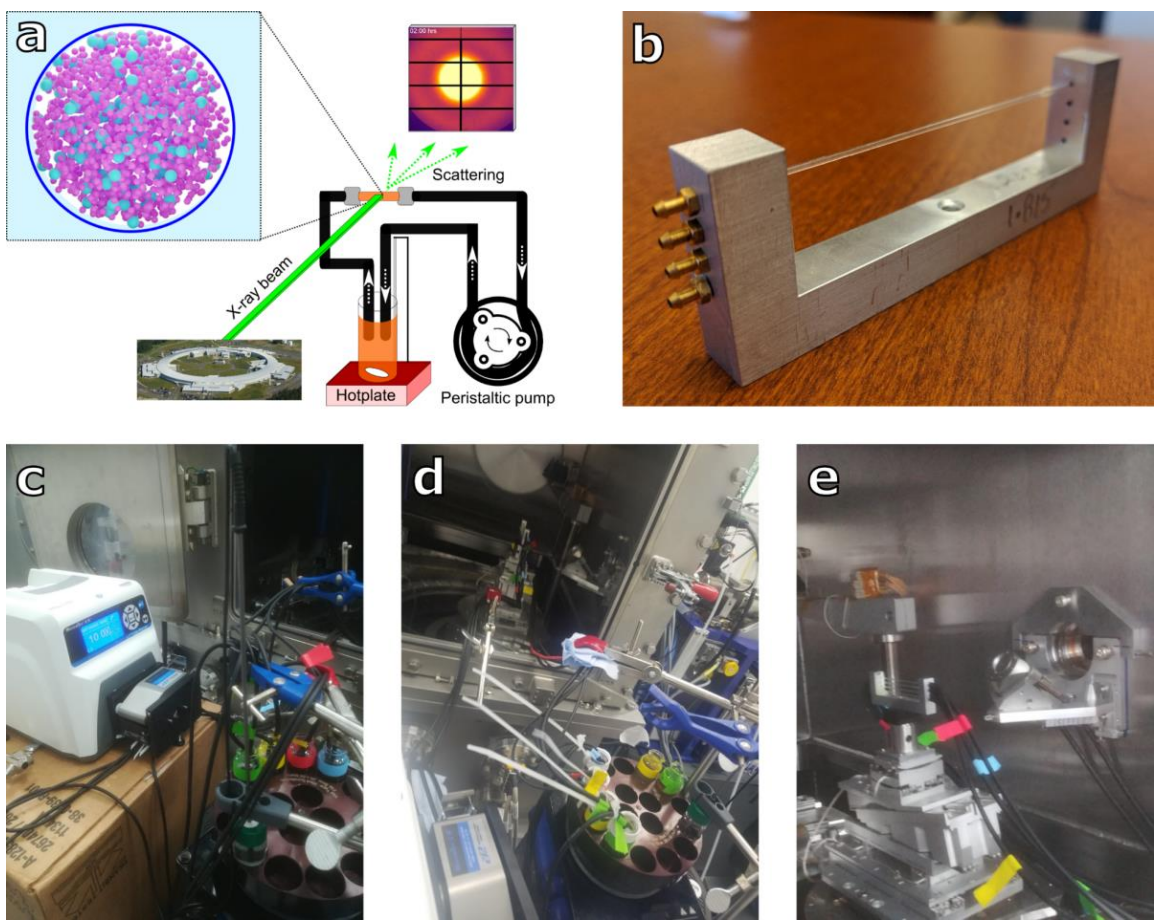

**Figure S4:** Schematic of the setup used to measure the kinetic SAXS patterns. **(a)** Pictorial representation of the setup: The emulsion moves via peristaltic flow in a closed loop that includes a quartz capillary tubing. The X-ray beam is aligned with the center of the tubing. As the emulsion evaporates from the thermo-stated vial, the scattering pattern changes. **(b)** Photograph of the sample holder used for the experiment. This holder was designed to support four stacked capillaries. The capillaries were glued to the aluminum holder using 5-minute epoxy (Devcon). Connection to the peristaltic tubing is ensured via barbed connectors. **(c-e)** Photographs of the complete setup. Four vials containing four distinct emulsions were secured on the hot plate. The Viton peristaltic tubing delivered the emulsion to the sample holder at a flow rate of 10 mL/minute.

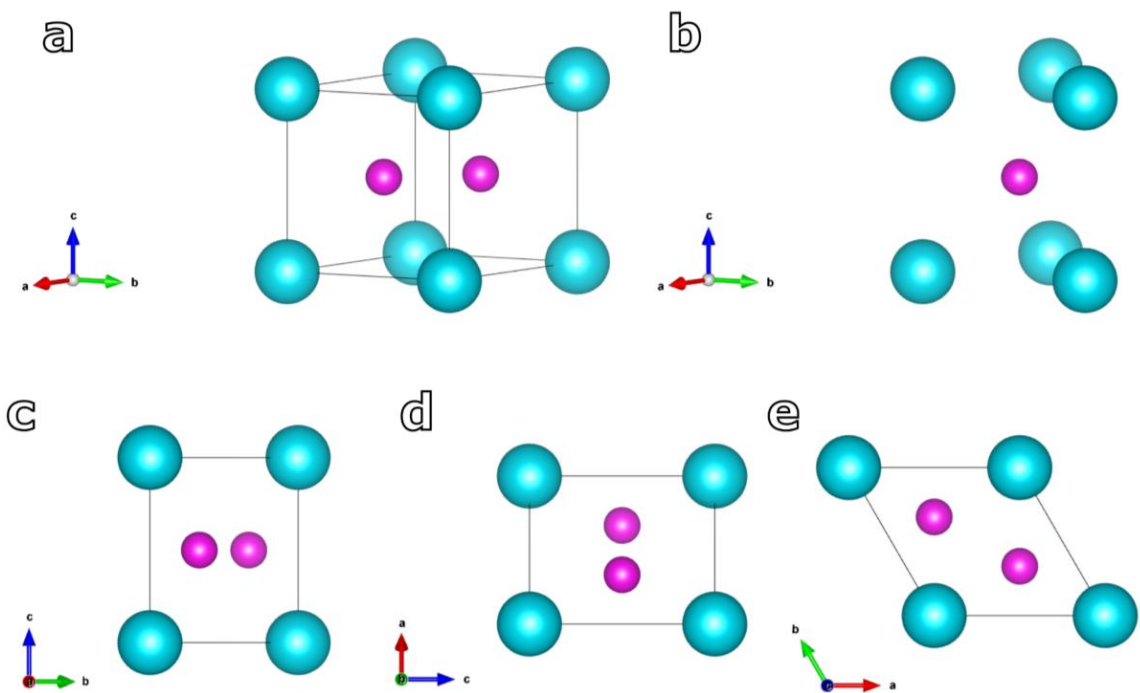

**Figure S5:** Unit cell of the crystal structure AlB$_2$. **(a)** The unit cell features 8 corner-sharing L particles and 2 S particles. **(b)** Each S particle occupies the trigonal prismatic voids left by 6 neighboring L particles. **(c)** Unit cell oriented with the *a* **(c)**, **(d)** *b*, and *c* **(e)** axis oriented orthogonal to the page.

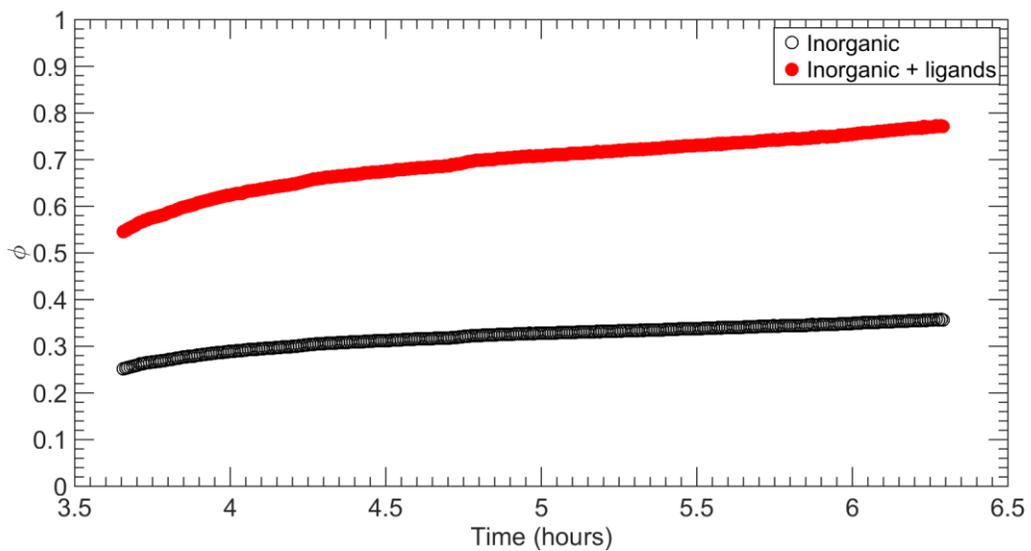

**Figure S6:** Kinetics of the volume fraction of the AlB$_2$ BNSLs nucleated from Fe$_3$O$_4$ (L) and PbS (S) nanocrystals at a NC number ratio of 1:2. Calculation details are described in the supplementary text.

S6

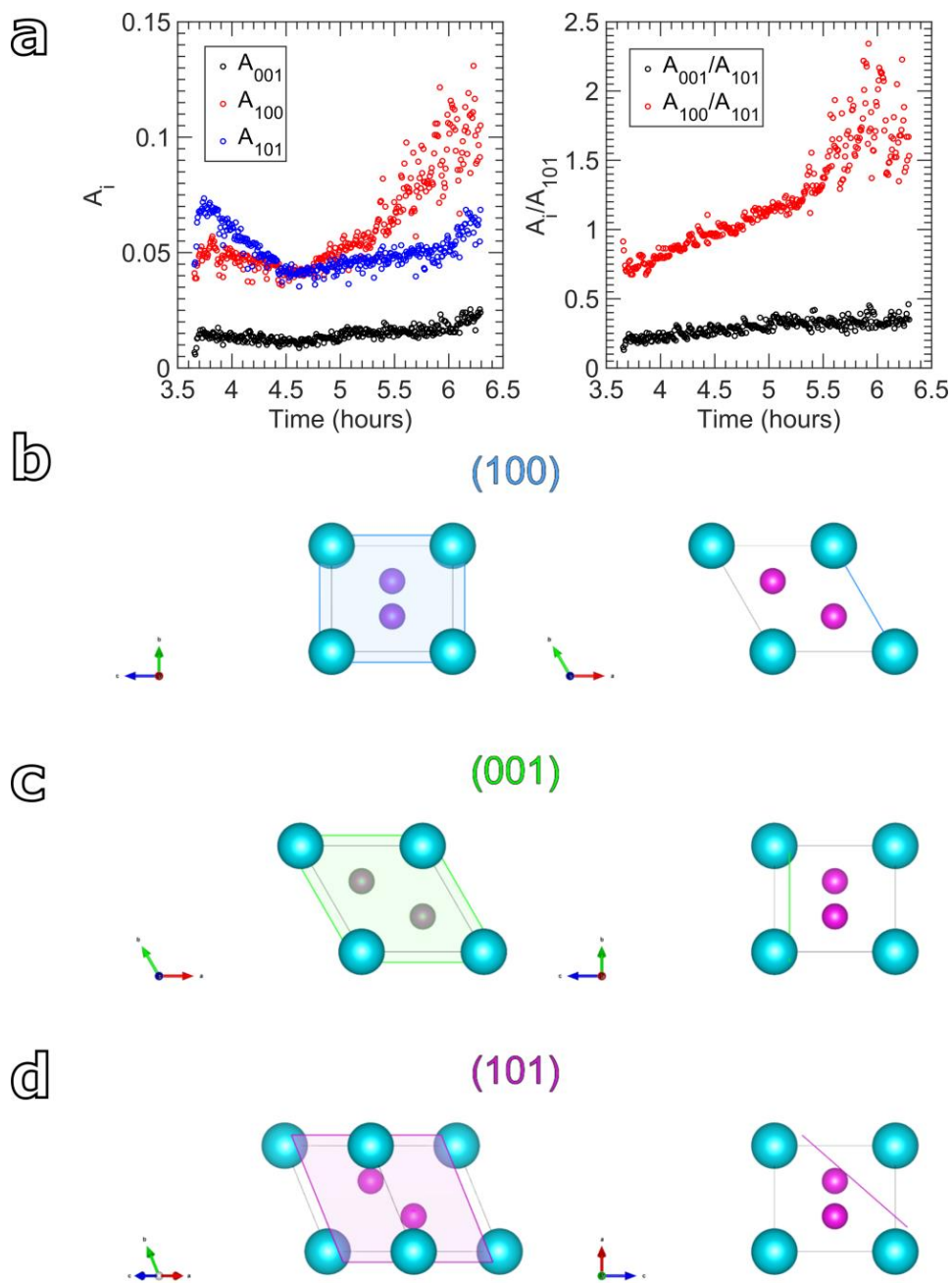

**Figure S7: (a)** Integrated intensities of 001, 100, and 101 reflections (left) and relative intensities of the 001 and 100 reflections relative to the 101 reflection during the growth of $AlB_2$ BNSLs nucleated from $Fe_3O_4$ (L) and PbS (S) nanocrystals at a NC number ratio of 1:2. **(b-d)** Highlights of the 100, 001, and 101 planes of the $AlB_2$ structure, respectively, for different orientations of the unit cell.

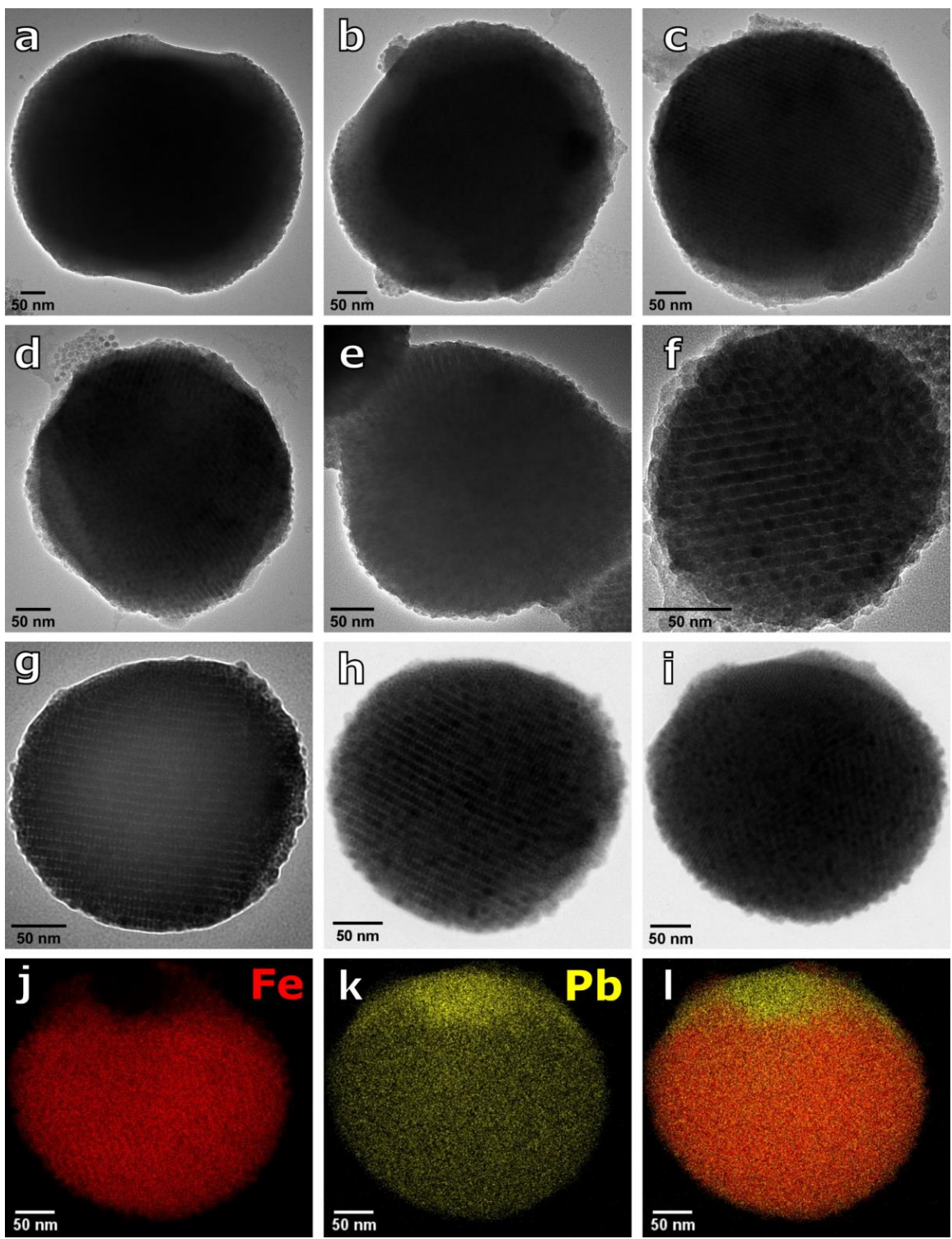

**Figure S8:** TEM **(a-g)**, STEM **(h-i)**, and EDS-STEM **(j-l)** micrographs of the AlB$_2$ BNSLs prepared with a NC number ratio of Fe$_3$O$_4$ : PbS#1 = 1 : 2. The fringes confirm the crystalline nature of the superstructures as suggested by *in situ* SAXS.

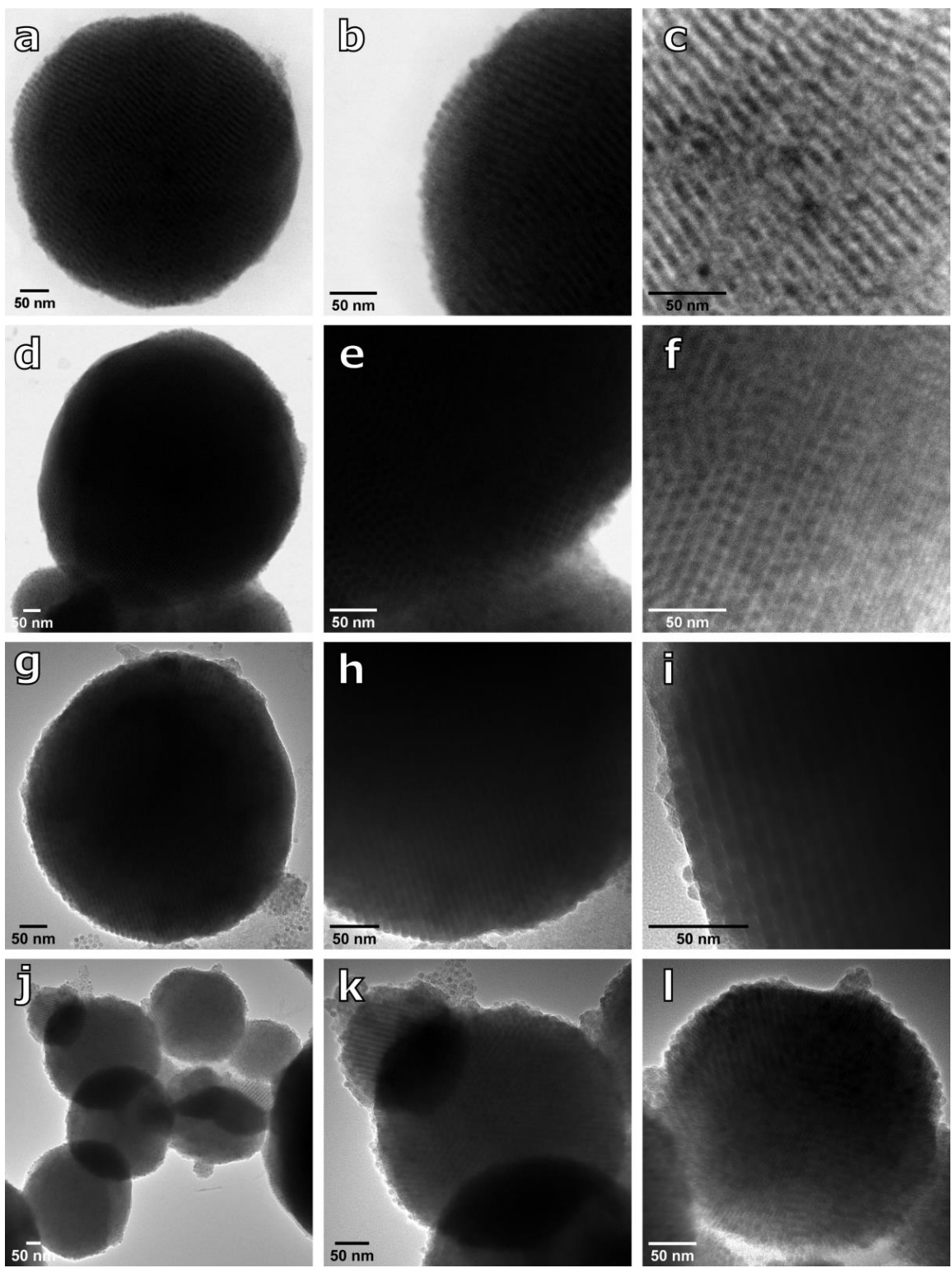

**Figure S9:** Additional STEM **(a-f)** and TEM **(g-l)** micrographs of the AlB$_2$ BNSLs prepared with a NC number ratio of Fe$_3$O$_4$ : PbS#1 = 1 : 2. Varying the magnification further confirms the crystallinity of superstructures of different sizes, both far from and close to the edge.

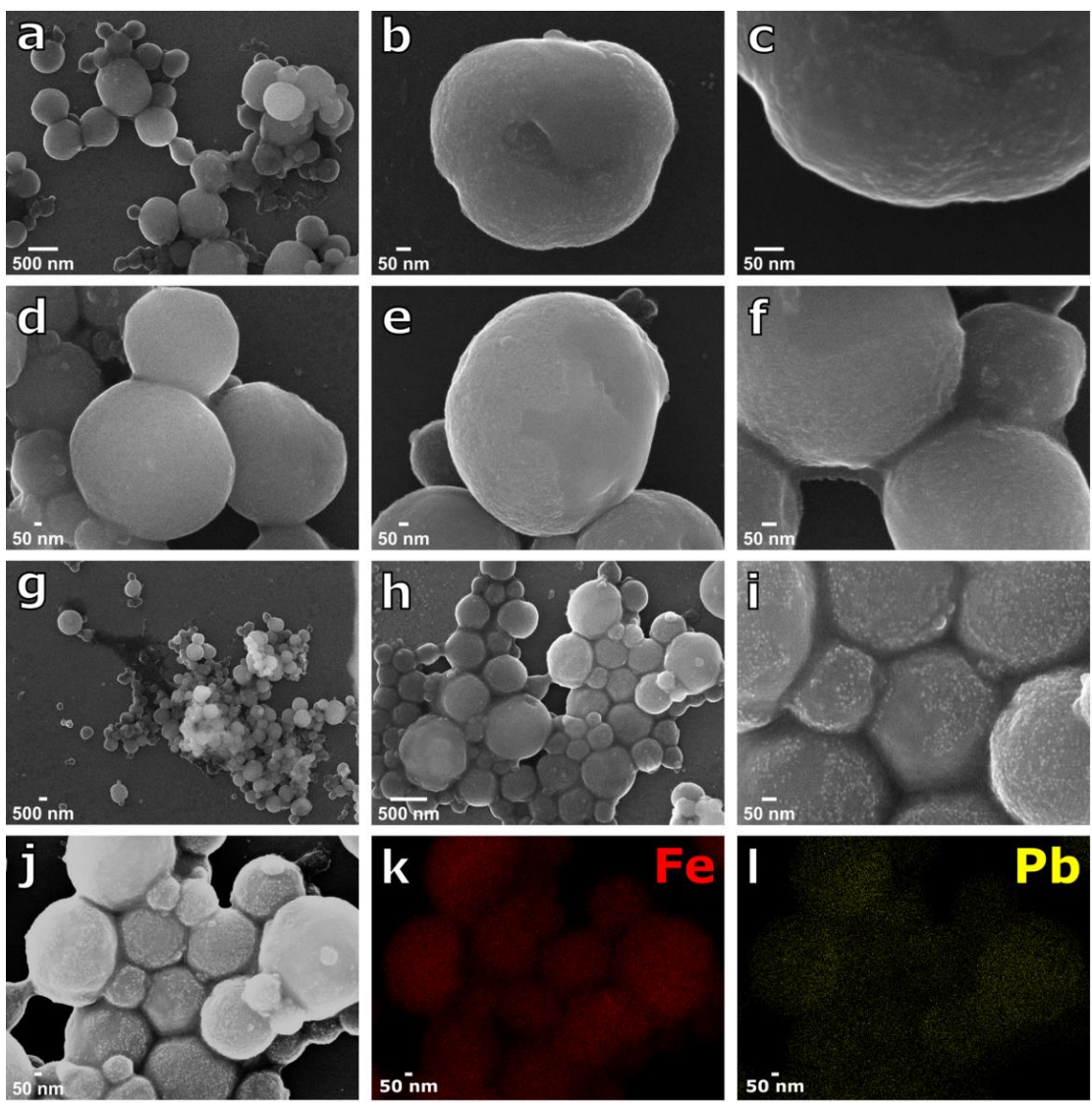

**Figure S10: (a-j)** SEM micrographs of the AlB$_2$ BNSLs prepared with a NC number ratio of Fe$_3$O$_4$ : PbS#1 = 1 : 2. The superstructures assume a globular shape imposed by the emulsion template. The occasional deviation from the spherical shape can be ascribed to the low surface tension imposed by the high surfactant concentration in the continuous aqueous phase. **(k-l)** EDS mapping confirms the uniform mixing of the two nanocrystal sizes.

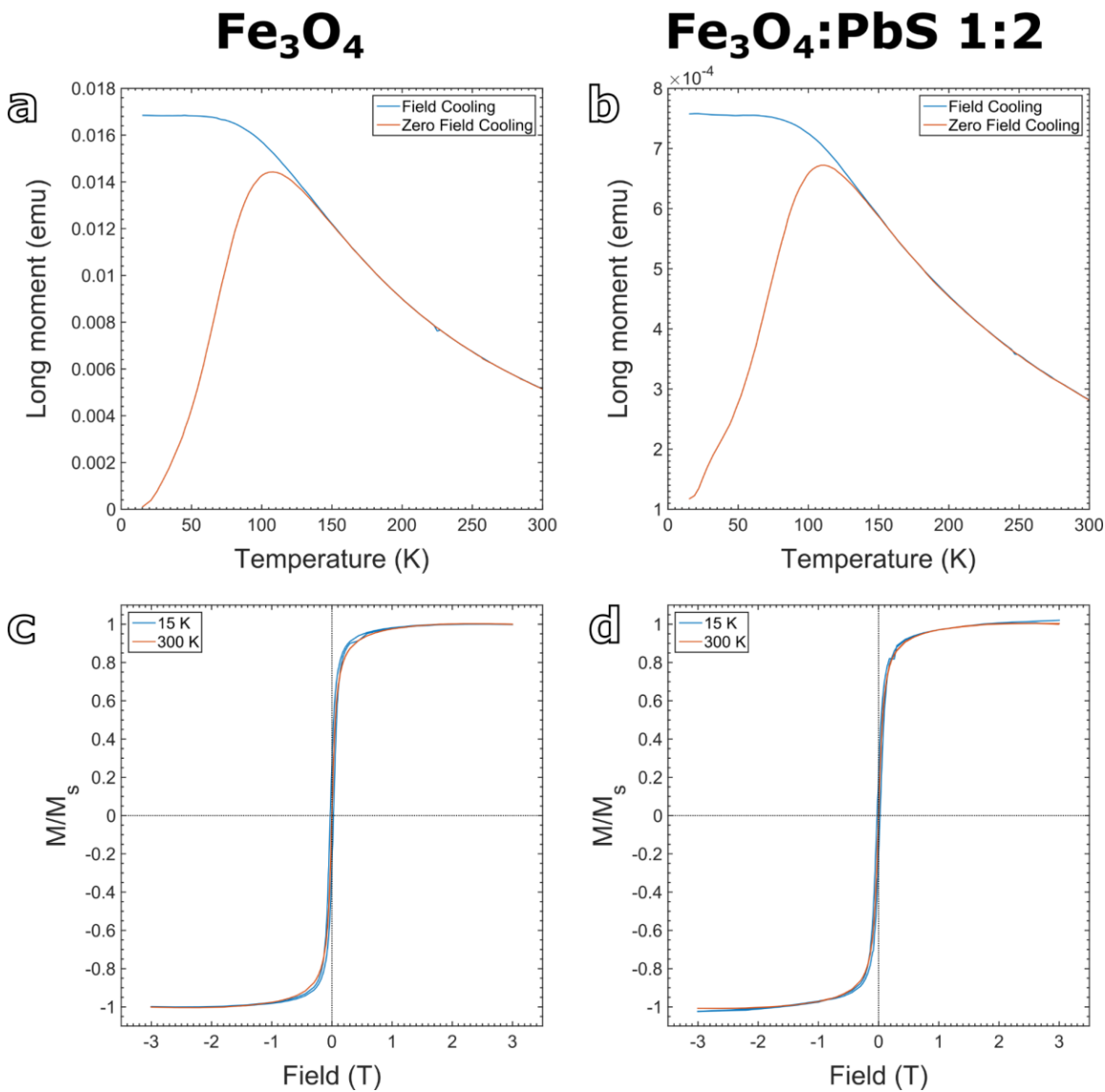

**Figure S11:** Results of the magnetic measurements performed on $Fe_3O_4$ **(a, c)** and $AlB_2$ BNSLs formed from $Fe_3O_4$ : PbS#1 NCs with a number ratio of 1 : 2 **(b, d)**.

| Sample | $Fe_3O_4$ | $Fe_3O_4$:PbS 1:2 |
|---|---|---|
| **Blocking temperature (K)** | 136 | 140 |
| **Coercivity (Oe)** | 240 | 240 |
| **Remanence ($10^{-4}$ emu)** | 83 | 16 |

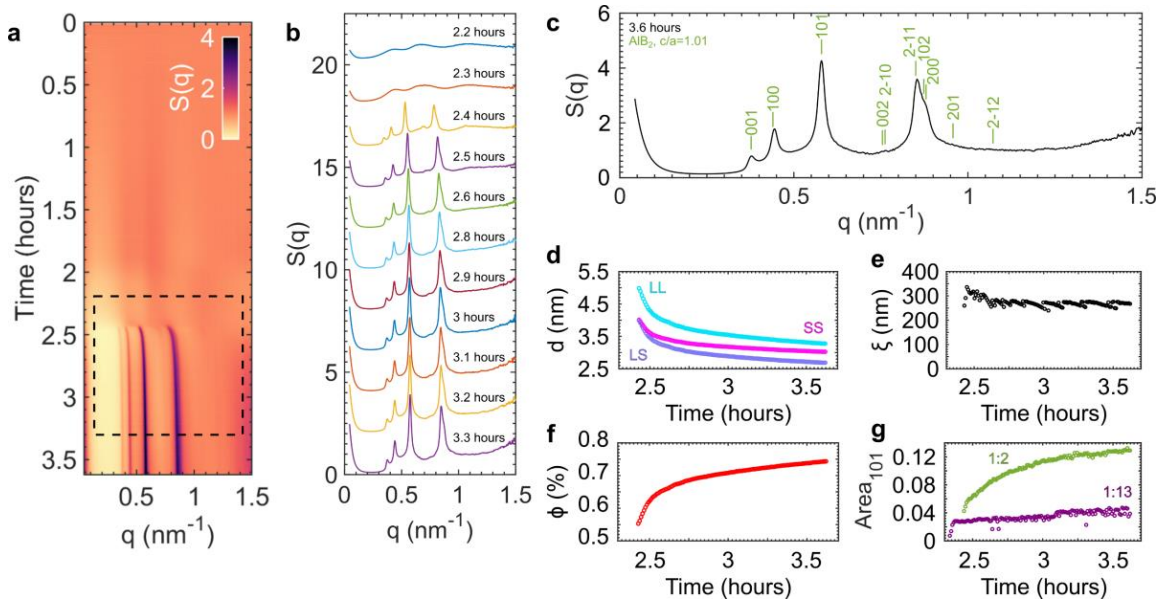

**Figure S12:** Formation of AlB$_2$ BNSLs nucleated from CdO (L) and PbS (S) NCs at a NC number ratio of 1 : 2. **(a-b)** Kinetic structure factor showing the emergence of diffraction peaks after 2.5 hours of drying. **(c)** Final structure factor confirming the AlB$_2$ crystal structure with c/a=1.01. **(d)** Surface-to-surface distance between L and S particles during lattice compression. **(e)** Average crystal size measured during assembly. **(f)** Effective volume fraction of the AlB$_2$ crystal measured during assembly. **(g)** Area of the 101 reflection used to extract the kinetic structural parameters for the two different NC number ratios investigated in the text.

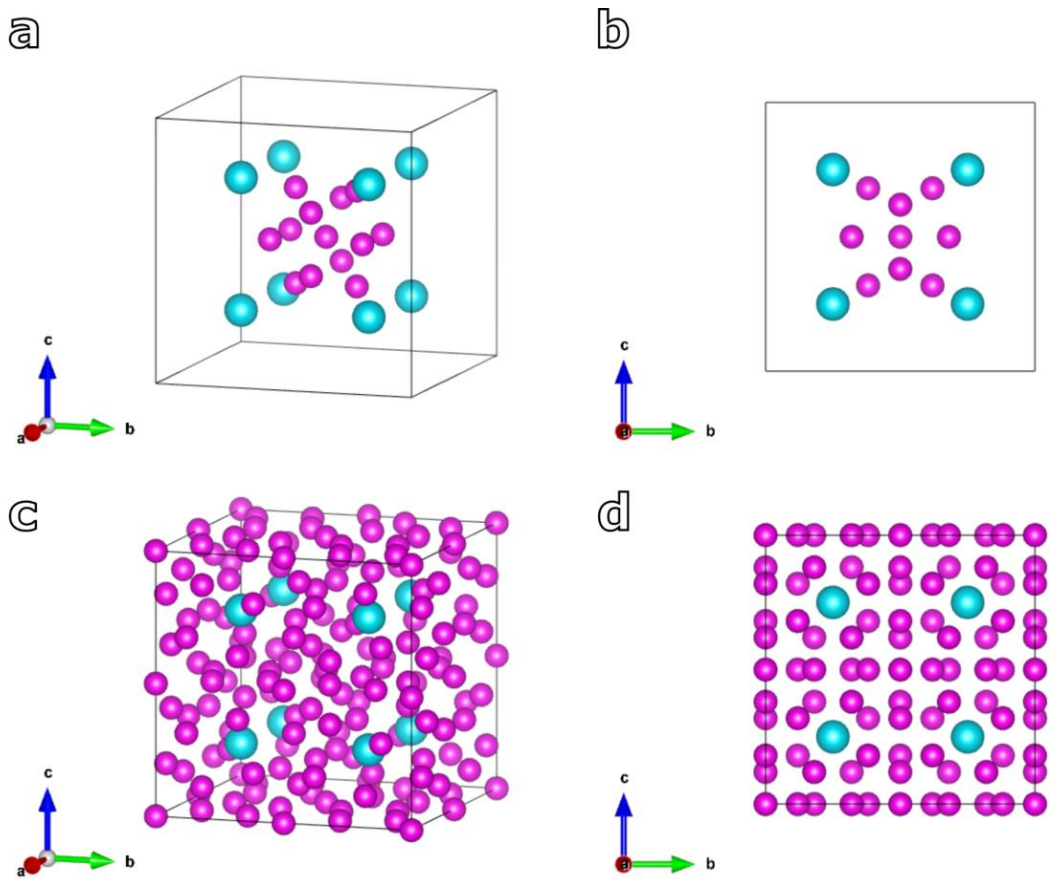

**Figure S13:** Unit cell of the crystal structure NaZn$_{13}$. **(a-b)** The simple-cubic sub-unit cell features 8 corner-sharing L particles and 13 S particles. **(c-d)** The unit cell of the structure is also cubic but with a side length twice that of the sub-unit cell.

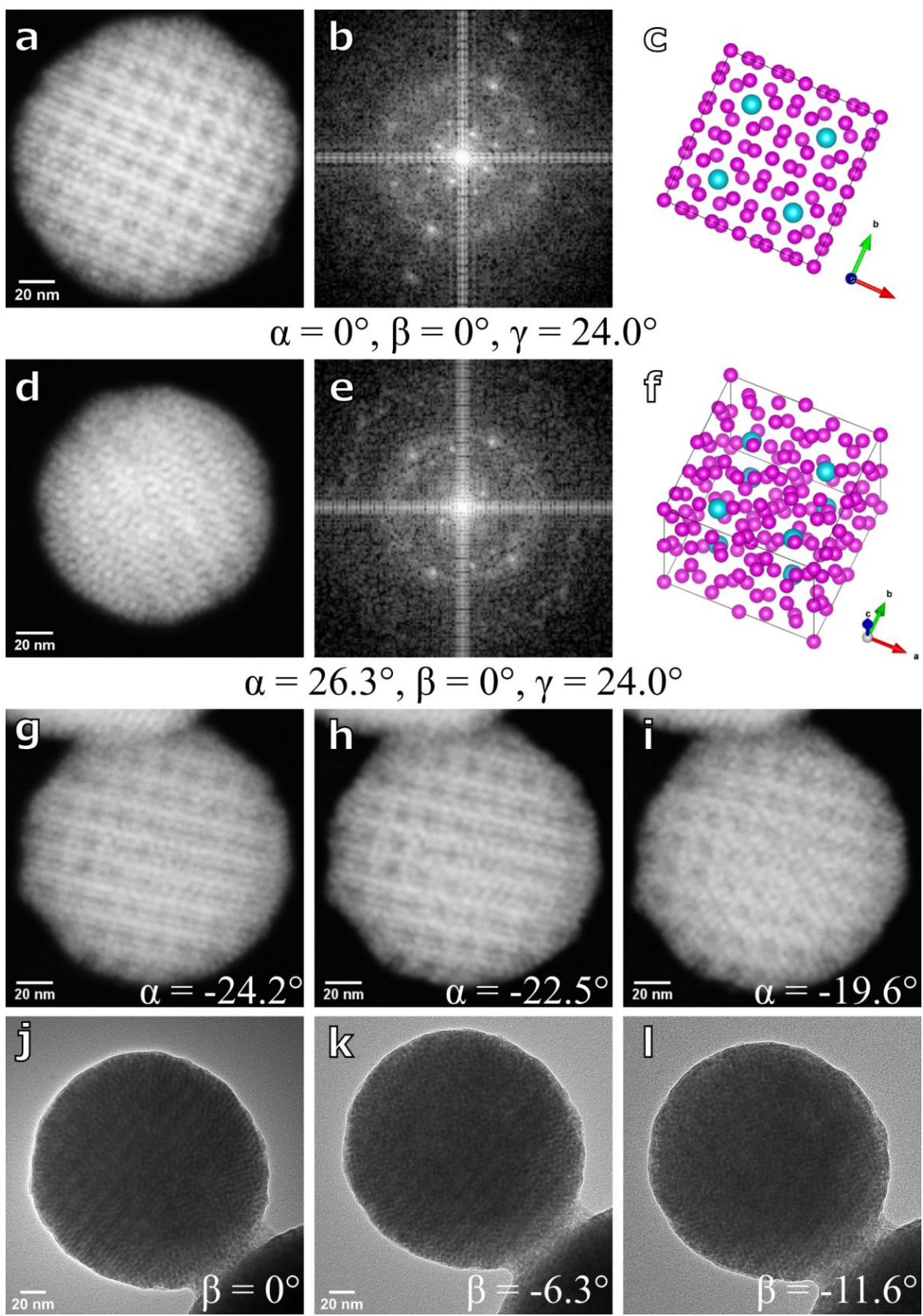

**Figure S14:** STEM **(a,d,g-i)** and TEM **(j-l)** micrographs illustrating the NaZn$_{13}$ BNSLs prepared with a NC number ratio of FICO : PbS#2 = 1 : 13 at different tilt angles. **(b, e)** represent the FFT image for **(a,d)**, while **(c,f)** show a proposed representation of the particle shown in (a, d).

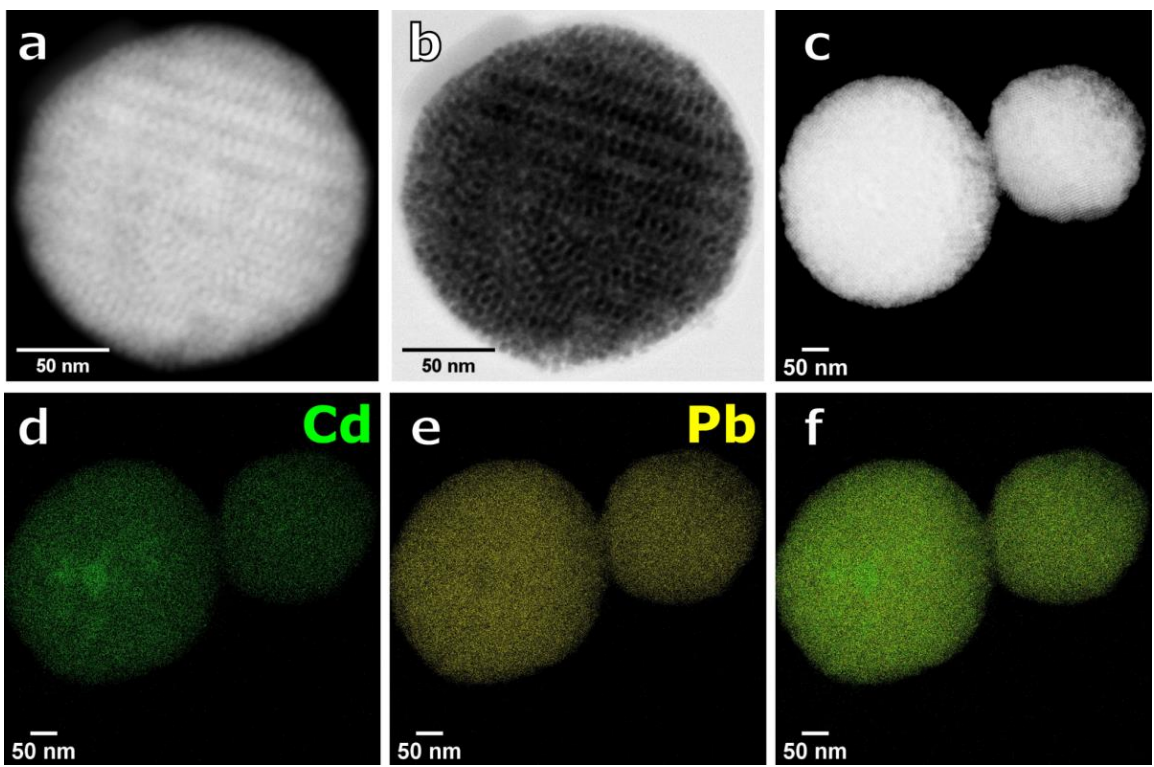

**Figure S15:** STEM **(a-c)** and EDS-STEM **(d-f)** micrographs of the NaZn$_{13}$ BNSLs prepared with a NC number ratio of FICO : PbS#2 = 1 : 13. (a) and (b) show bright- and dark-field STEM micrographs of the same BNSL. (c) shows the dark-field STEM micrograph of the BNSLs investigated by EDS-STEM (d-f).

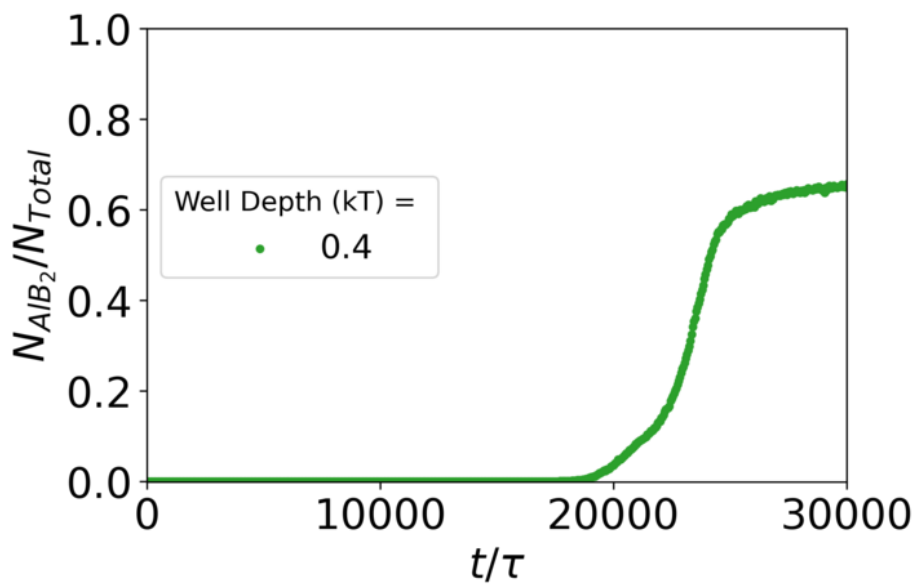

**Figure S16:** Self-assembly of AlB$_2$ with nonlinear scaling of the well-depth with core size. Here we repeated the self-assembly simulations shown in Figure 3e, except we set the *m* coefficient of the Mie potential to 49 for interactions between small particles and to 37 for interactions between small and large particles, while keeping *m* = 25 for interactions between large NCs. Changing the *m* coefficient in this manner significantly reduces the range of the attractive well between small NCs and between small NCs and large NCs. The simulations were run similarly to those in Figure 3e, except we needed to increase the densities to observe self-assembly. The specific densities we compressed over were $\rho\sigma^3 = 2.23$ to $\rho\sigma^3 = 2.43$. We find that AlB$_2$ still self-assembles, which implies that its self-assembly is relatively insensitive to the scaling the of attractive well with NC size (provided that the wells are still fairly narrow)."

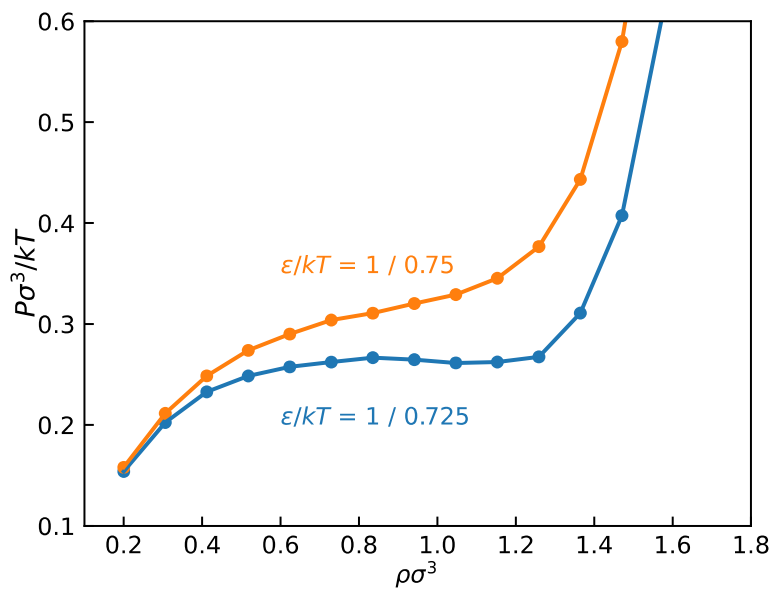

**Figure S17:** Our estimation of the critical point using pressure-volume data. The downward slope between a few points at $\varepsilon/kT = 1/0.725$ indicates that vapor-liquid separation occurs while the lack of a downward slope for $\varepsilon/kT = 1/0.75$ indicates the lack of phase separation. We thus estimated the critical point to be $\varepsilon/kT = 1/0.738$. Note that this estimate is not used in any of our free energy calculations, and thus the estimate's uncertainty does not propagate into those calculations. We collected the pressure-volume data by simulating 8,000 particles for $16,000\tau$ at a stoichiometry of 1:2.

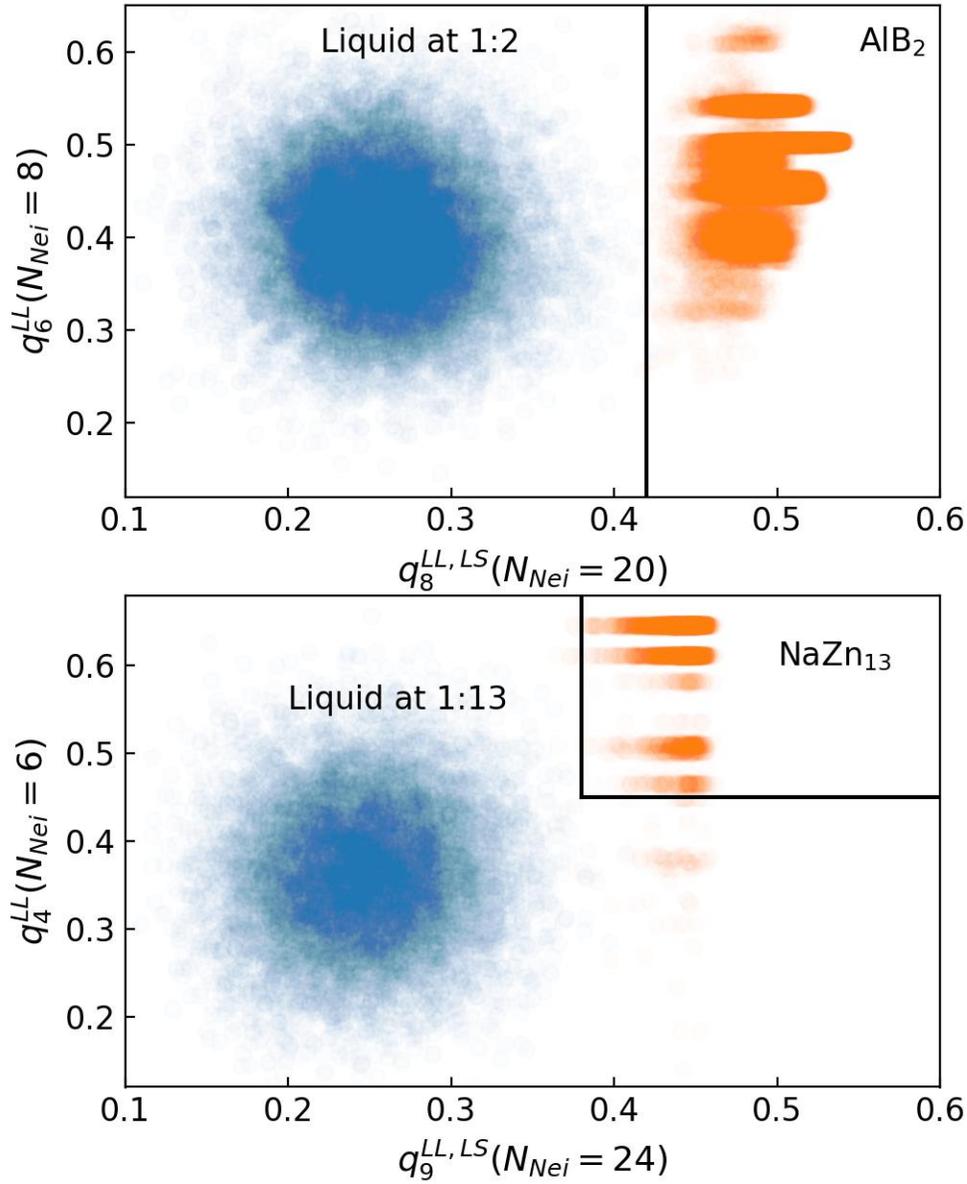

**Figure S18:** The order parameters we used to distinguish between liquid and solid particles. Large particles in the AlB$_2$ phase can be distinguished using the $q_8$ using the closest 20 particles (large or small). We also computed the $q_6$ of the nearest 8 large particles but do not use it identify particles as being in the AlB$_2$ phase. Large particles in the NaZn$_{13}$ phase can be distinguished using the $q_4$ using the closest 6 large particles and the $q_9$ of the closest 24 particles (large or small). The number of neighbors was chosen to match that of particles in the perfect crystal. The results for the liquid phase were taken from the first few frames of our self-assembly simulations shown in Figure 4. The results for the solid phases were gathered at $P\sigma^3/\varepsilon = 1$ and $\varepsilon/kT = 2.5$. To account for the fact that our self-assembly results tend to have many defects, we removed about 10 % of the particles from each simulation frame of the solid phases before computing the order parameters. The specific bound used to distinguish AlB$_2$ is $q_8^{LL,LS} > 0.42$. The specific bounds used to distinguish NaZn$_{13}$ are $q_9^{LL,LS} > 0.38$ and $q_4^{LL} > 0.45$.

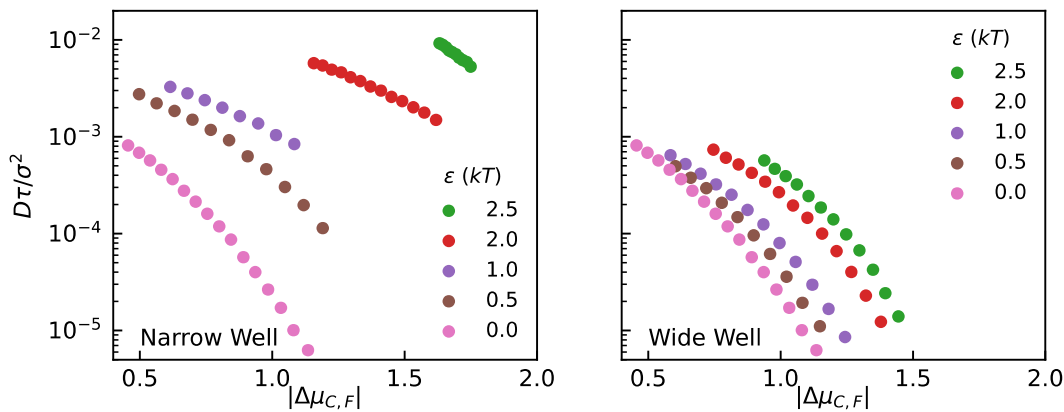

**Figure S19:** The diffusion constant versus the chemical potential driving forces for every pair potential we attempted self-assembly with in Figure 3. The chemical potential driving forces are computed as $|\Delta\mu_{C,F}| = \left|\hat{G}_{AlB_2} - \frac{\mu_L}{3} - \frac{2\mu_S}{3}\right|$, where the chemical potentials $\mu$ and the specific Gibb's free energy $\hat{G}$ are computed in the same manner used to generate the phase diagrams in Figure 3. Nucleation should be most favorable if a high $|\Delta\mu_{C,F}|$ can be achieved with highly mobile particles, which corresponds to the upper right corner of the graphs. We see a trend towards that corner with increasing well depth, with the trend being much stronger for the narrow well system.

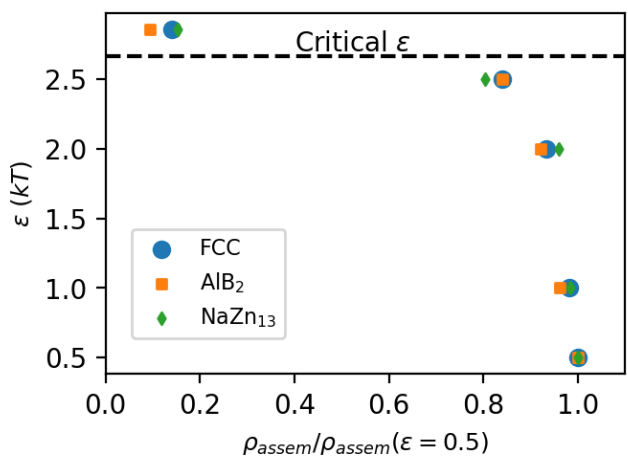

**Figure S20**: The densities at which different crystals begin to self-assemble for NCs interacting via narrow wells at different well depths. We normalized the densities relative to their values at $\varepsilon = 0.5\ kT$. Our results for FCC were obtained in a single component system of the large particles; our results for $AlB_2$ were obtained at a number ratio of 1:2; and our results for $NaZn_{13}$ were obtained at a number ratio of 1:13. Simulations were conducted through slow compression. The beginning of self-assembly was detected by examining the density at which a small number of large NCs (60 for FCC, 30 for $AlB_2$, and 10 for $NaZn_{13}$) could be labelled crystalline. Densities are normalized with respect to the density for which they self-assemble at a well depth of 0.5 $kT$. For $AlB_2$ and $NaZn_{13}$, particles were labelled as crystalline using the order parameters described in Figure S17. For FCC, particles were labelled as crystalline if the Steinhardt order parameters $q^6 > 0.42$ and $q^{10} < 0.14$ when they are computed over the first 12 neighbors of each particle. We estimated the critical density at a stoichiometry of 1:2; our results indicate that it does not vary substantially at 1:0 or 1:13. We show how the critical density is estimated in Figure S20. In all cases, the density at which self-assembly occurs decreases with increasing well depth, particularly above the critical well depth.

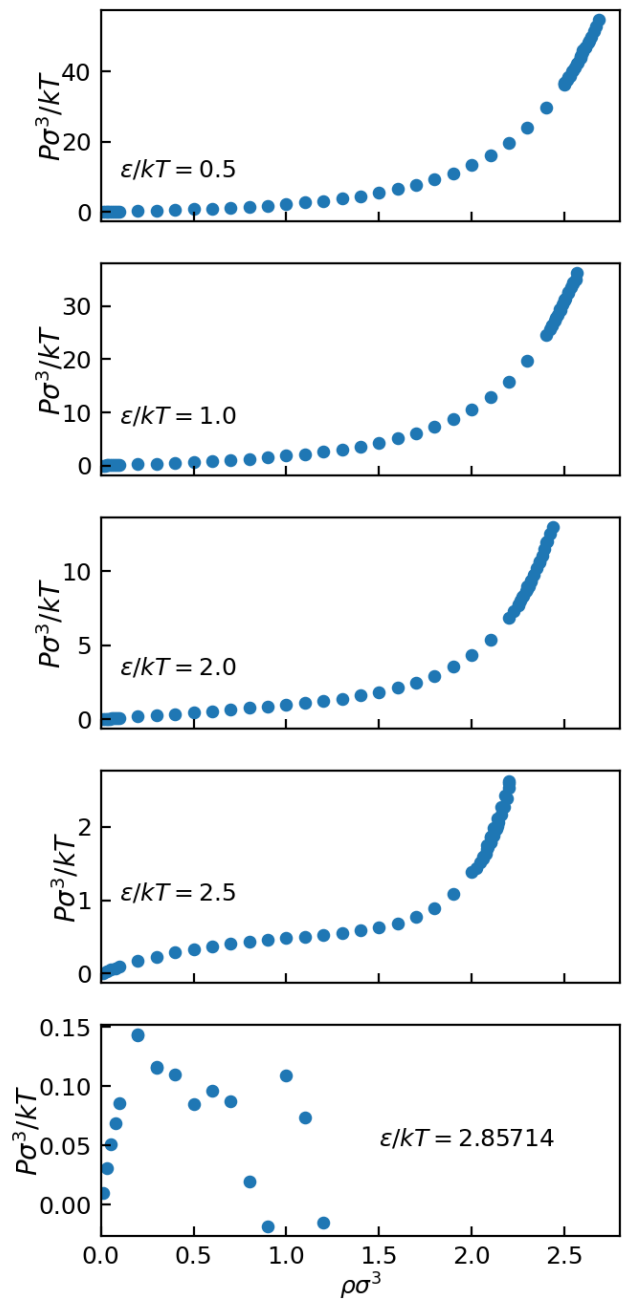

**Figure S21:** Pressure-volume data for the narrow well at the 4 different well depths examined in the text. The discontinuities present at $\varepsilon/kT = 2.85714$ indicate that the critical well depth is in between 2.5 and 2.85714 $kT$. In Figure S20, we take it to be 2.68 $kT$, which is the midpoint of these two values. We note that any inaccuracy in our estimation of the critical point does not influence the accuracy of any other calculation we perform. Here we only show the results for a number ratio of 1:2, but our results in Figure S19 that the critical well depth does not vary substantially at 1:0 or 1:13.

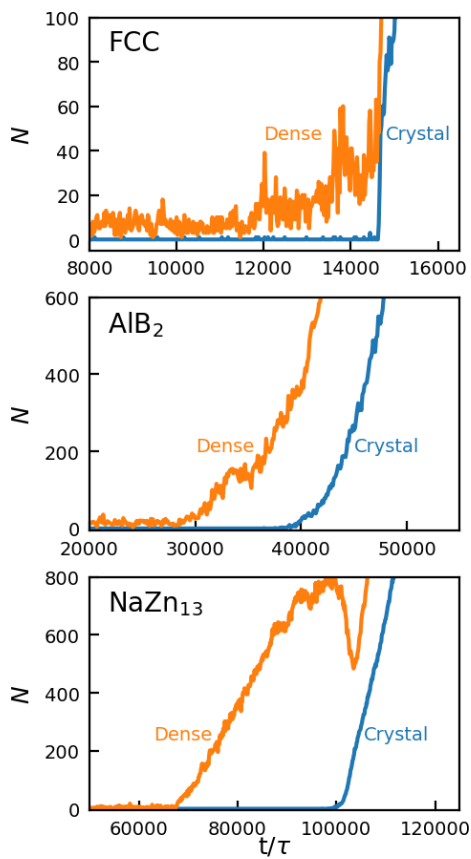

**Figure S22.** Two-step self-assembly with NCs interacting via a narrow well at well depths greater than the critical well depth. Our results for FCC were obtained in a single component system of the large particles; our results for $AlB_2$ were obtained at a number ratio of 1:2; and our results for $NaZn_{13}$ were obtained at a number ratio of 1:13. We used a well depth of 2.85714 $kT$. We show the number $N$ of large particles in these systems identified as locally dense or crystalline. For $AlB_2$ and $NaZn_{13}$, particles were labelled as crystalline using the order parameters described in Figure S17. For FCC, particles were labelled as crystalline if the Steinhardt order parameters $q^6 > 0.42$ and $q^{10} < 0.14$ when they are computed over the first 12 neighbors of each particle. See methodology for more details on how we compute the local density. The simulations for each crystal were run at different densities because different crystal self-assemble at different densities. Specifically, we obtained the results for FCC by slowly compressing from $0.11/\sigma^3$ to $0.15/\sigma^3$, $AlB_2$ from $0.21/\sigma^3$ to $0.24/\sigma^3$, and $NaZn_{13}$ from $0.35/\sigma^3$ to $0.45/\sigma^3$. We show the initial rise in $N$ to emphasize whether locally dense or crystalline particle appear first. In every case, we observe a rise in the number of locally dense NCs before observing a rise in the number of crystalline particles. The initial rise in locally dense particles is more dramatic for $AlB_2$ and $NaZn_{13}$, indicating that the locally dense particles take a longer time to rearrange into the crystal.

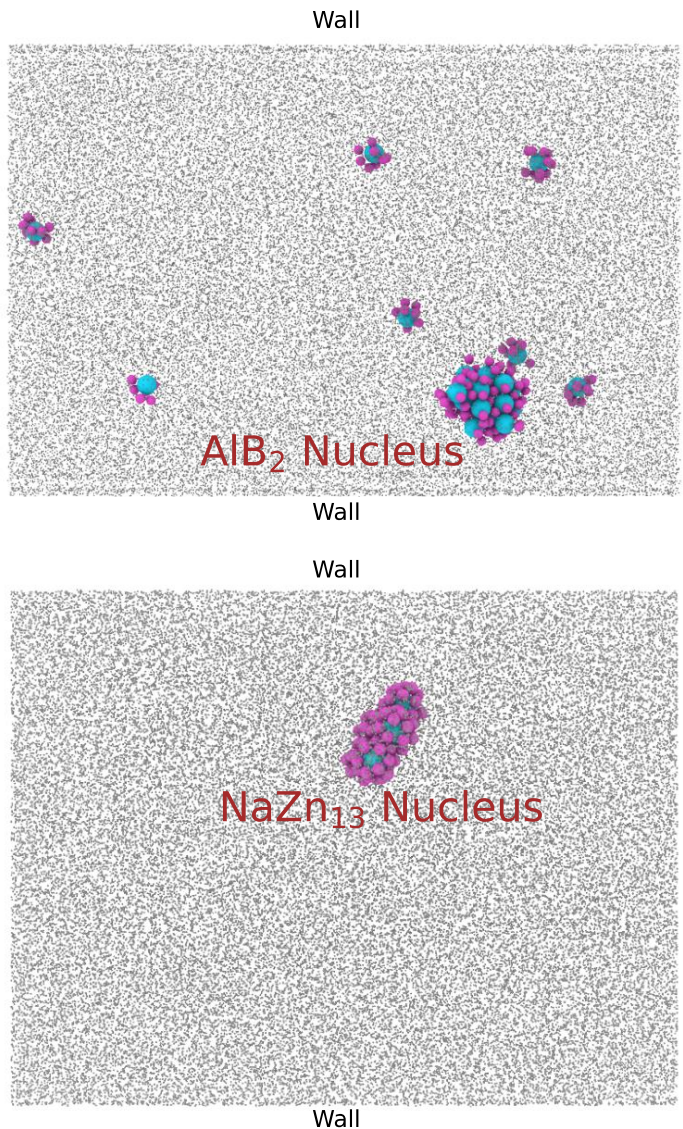

**Figure S23:** BNSL formation with flat walls. We show any large NC identified as crystalline as blue and any small particle neighboring a crystalline large NC as pink; all other NCs are reduced in size and colored grey. The simulations were run under conditions close to those shown in Figure 4. The $AlB_2$ nucleus forms at a stoichiometry of 1:2, while the $NaZn_{13}$ nucleus forms at a stoichiometry of 1:13. In no case does the nucleation begin on the wall.

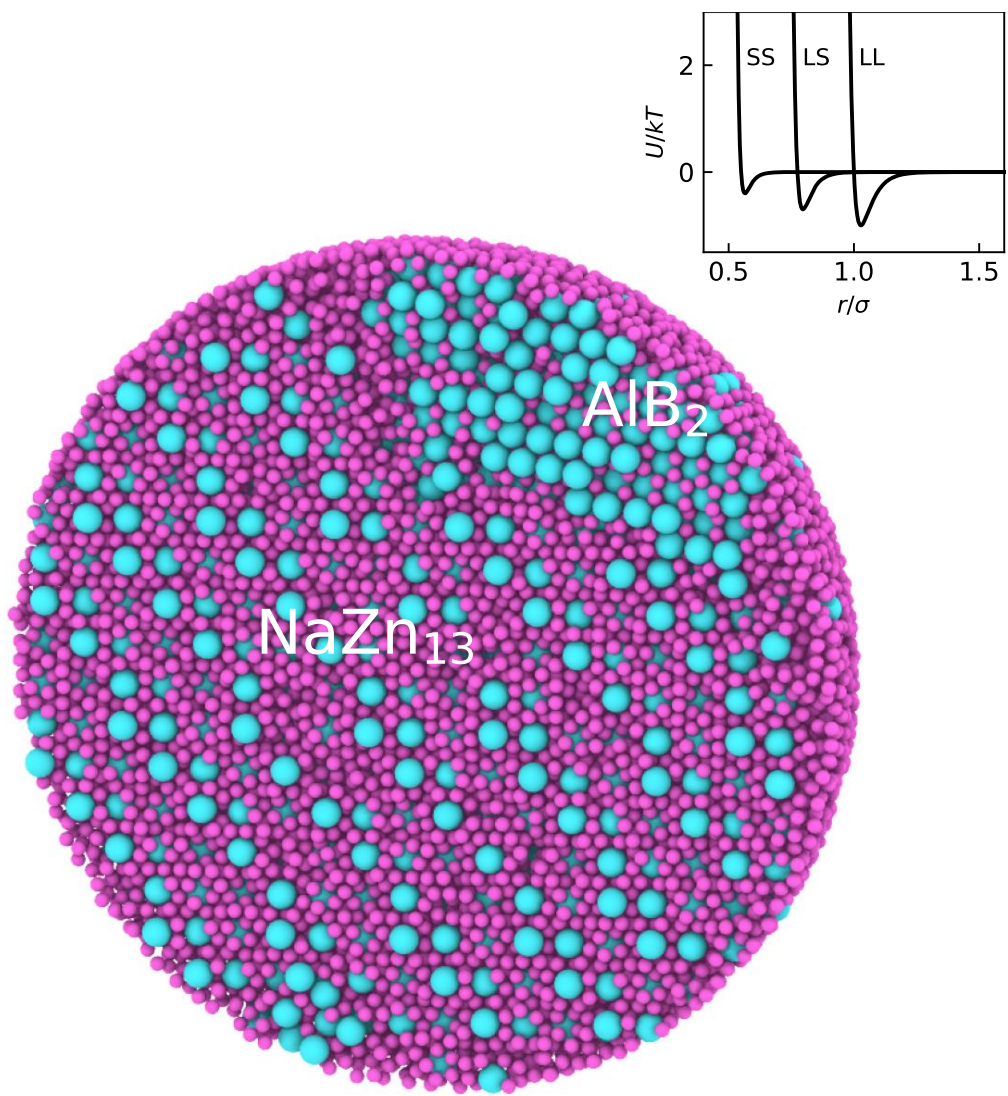

**Figure S24:** Crystal formation with unequal attractive wells. As shown in the plot, the well-depth is 2.5 kT between large particles, 1.75 kT between large and small particles, and 1 kT between small particles. With this pair potential we obtain a small region of AlB$_2$ alongside a larger region of NaZn$_{13}$ at a stoichiometry of 1:13. We did not use polydisperse particles for this simulation because polydispersity slows equilibration.

| *m* | *ε (kT)* | **Densities ($\rho\sigma^3$)** |
|---|---|---|
| 25 | 0.0 | 2.60 – 2.70 |
| 25 | 0.5 | 2.60 – 2.70 |
| 25 | 1.0 | 2.52 – 2.62 |
| 25 | 2.0 | 2.37 – 2.47 |
| 25 | 2.5 | 2.18 – 2.28 |
| 6 | 0.0 | 2.60 – 2.70 |
| 6 | 0.5 | 2.61 – 2.71 |
| 6 | 1.0 | 2.54 – 2.64 |
| 6 | 2.0 | 2.44 – 2.54 |
| 6 | 2.5 | 2.41 – 2.51 |

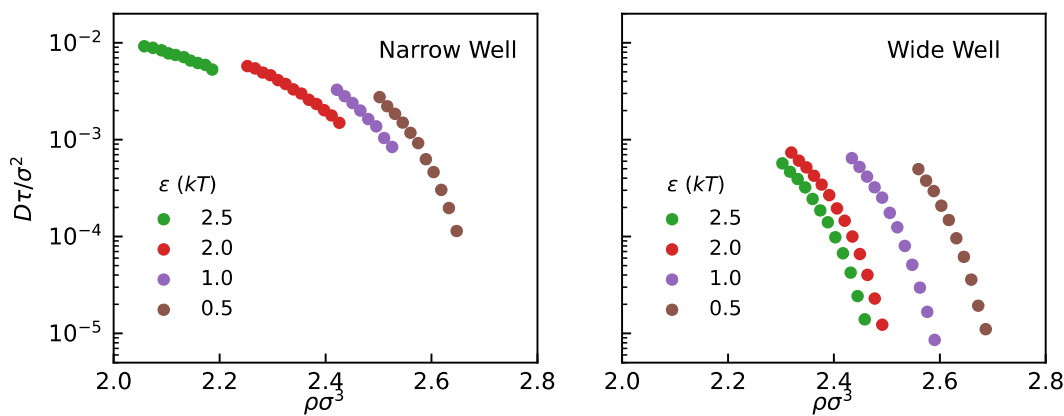

**Figure S25:** The range of densities we compressed over in Figure 3e-f, and the diffusion constants (*D*) of the large particles as a function of density for different well-depths and well-widths. We computed *D* from simulations of 27,000 particles at a NC number ratio of 1:2. We could collect data up until crystallization began to occur; thus, we could go to higher densities, and lower diffusion coefficients, for the wide well. Note that we varied well depth by changing the temperature instead of explicitly $\varepsilon$. We expect that changing $\varepsilon$ instead would change the diffusion coefficients by less than an order of magnitude.

# NANOCRYSTAL SYNTHESIS

**PbS NCs:** PbS nanocrystals (NCs) were synthesized by following the method developed by Hines and Scholes[1] and optimized for monodispersity by Voznyy *et al.*[2]

**Materials:** 1-Octadecene (ODE, technical grade, Acros Organics), oleic acid (OA, technical grade, Sigma Aldrich), oleylamine (technical grade, Sigma Aldrich), PbO (99.9+% trace metals basis, Acros Organics), $PbCl_2$ (99.999%, Alfa Aesar), hexamethyldisilathiane (($TMS)_2S$, synthesis grade, Sigma Aldrich), hexanes (certified ACS, Fisher), and acetone (certified ACS, Fisher).

**Precursor preparation:** In a 250 mL round-bottom flask, 9.0 g of PbO, 30 mL of oleic acid, and 60 mL of 1-octadecene were mixed and stirred at 110 °C overnight under a vacuum of $10^{-1}$ Torr. Meanwhile, in the glove box, an amount of $(TMS)_2S$ was added to dried 1-octadecene. Exact amounts are specified below in the table below. Separately, a 0.3 mM solution of $PbCl_2$ in oleylamine was prepared by stirring at 120 °C under vacuum.

**NC synthesis:** An amount of lead oleate precursor was added to a 100 mL three-neck round-bottom flask equipped with a thermocouple adapter and charged with 15 mL of ODE. The flask was degassed at 100 °C for one hour. After changing the atmosphere to nitrogen, the flask was allowed to cool to ∼50 °C, followed by slow heating to the injection temperature. Once the target temperature was reached the $(TMS)_2S$/ODE solution was swiftly injected into the flask by using a 22 mL plastic syringe equipped with a 16 G needle. After annealing for a specified time, the heating mantle was turned off and the flask was allowed to cool naturally. 1 mL of the $PbCl_2$/oleylamine solution was injected into the flask when the mixture reached 60 °C. The flask was then allowed to cool naturally to 30 °C. The synthetic mixture was split in 50 mL centrifuge tubes, and acetone was added to achieve a volume ratio of 1:1. The cloudy dispersion was centrifuged at 8000 g for 3 minutes and the supernatant was discarded. The NCs were dissolved in toluene and the precipitation with acetone was repeated twice. The NCs were redispersed in 10 mL of toluene and filtered by using a 200 nm PTFE or PVDF syringe filter. The NC concentration was determined by spectrophotometry using a sizing curve reported in the literature.[3]

Note: We noticed that the efficiency of the first round of separation of the NCs depended somewhat on the batch of ODE used. Occasionally, the NCs would form an oily phase at the bottom of the centrifuge tube. In this case, we found that adding 5 mL of toluene to 15 mL of reaction mixture and mixing prior to the addition of acetone dramatically improved separation.

**Size separation:** No size separation was applied to these NCs.

| | | |
|---|---|---|
| Volume $(TMS)_2S$ | 206 μL | 240 μL |
| Volume ODE | 9.6 mL | 9.0 mL |
| Volume $Pb(OA)_2$ | 17.4 mL | 18.0 mL |
| Injection temperature | 110 °C | 126 °C |
| Annealing time | 5 minutes | 0 minutes |
| NC diameter as determined from SAXS Format: Average ± Standard deviation | 4.5 ± 0.4 nm | 6.4 ± 0.6 nm |

**Fe₃O₄ NCs:** Spherical iron oxide NCs approximately 10 nm in diameter were synthesized *via* thermal decomposition of an iron-oleate precursor which acts as a growth source. The synthesis was adapted from Park *et al.*[4] and modified from Yun *et al.*[5] The iron oleate precursor is prepared in-house by reacting iron chloride and sodium-oleate. The synthesis is relatively robust and repeatable and is tailored to yield a reasonably monodisperse sample directly from synthesis. A careful size selection process is used to further improve the monodispersity post-synthesis. The Fe₃O₄ NCs that are produced have a cubic spinel structure and display superparamagnetic behavior. It has been reported that the surface can become slightly oxidized to γ-Fe₂O₃ if exposed to air over time.[4] The NC surface is passivated with oleic acid (OA) ligands from synthesis.

**Materials:** Iron(III) chloride hexahydrate (FeCl₃ · 6H₂O, 97%), oleic acid (technical grade, 90%), and sodium oleate (>97%) were purchased from Sigma Aldrich. 1-octadecene (technical grade, 90%) was purchased from Acros Organics.

**Precursor preparation:** The first step is to prepare the iron oleate precursor. 21.6 g of iron(III) chloride hexahydrate, 73.0 g of sodium oleate, 80 mL of DI water, 80 mL of ethanol, and 160 mL of hexane were added to a 500 mL flask. The mixture was heated to 60°C while stirring with a stir bar and refluxed for 4 hours. The dark red iron oleate precursor is then washed with DI water three times in a conical separation flask. In the first step, ~200 mL of DI water is added, and the solution is vigorously mixed, then let to sit and phase separate, removing the denser phase from the bottom of the flask and keeping the dark red-brown phase. This process is repeated twice with fresh DI water, and ~50 to 100 mL of hexane is added during the second washing step to help drive phase separation. The water was removed as best as possible from the separation flask, and the precursor was transferred to a glass container and dried in a vacuum oven overnight at 80°C to form a viscous wax-like solid.

**NC synthesis:** The NCs were synthesized by mixing 7.2 g of the iron oleate precursor, 1.25 mL oleic acid, and 20 mL of 1-octadecene in a three-neck flask. The flask was placed under vacuum and the mixture was heated to 110 °C for 60 minutes, then placed under nitrogen and quickly heated to 310 °C at a rate of ~3°C/min and allowed to react at this temperature for 30 minutes. The heating mantle was removed, and the reaction vessel was cooled naturally in air to ~160°C, then quenched to room temperature using a water bath. The reaction material was purified using three washing steps. In step one, the reaction contents were divided evenly into four 50 mL centrifuge tubes (~5 mL of reaction material in each) and 5 mL of toluene was added to each tube. 40 mL of acetone was added as the antisolvent to make a 1:4 sample:acetone mixture, and the tubes were centrifuged at 9000 g for 5 minutes. The supernatant was discarded, 5 mL of toluene was added to each to redisperse the precipitated material, and the material from four tubes was combined into two tubes, leaving 10 mL of the NC dispersion in each. 10 mL of ethanol and 30 mL of isopropyl alcohol were added to each tube to make a 1:1:3 sample:ethanol:IPA mixture, which was centrifuged at 9000 g for 5 minutes. Again, the supernatant was discarded, and the NC precipitate was redispersed in 5 mL of toluene for each. The contents were combined into one tube (10 mL of NC dispersion) and 100 µl of oleic acid was added and mixed into the sample using a vortex mixer and mild sonication. In the final washing step, 40 mL of acetone was added to make a 1:4 sample:acetone mixture which was centrifuged at 8000 g for 5 minutes. The supernatant was discarded, and the precipitate was dried under vacuum into a solid pellet. The final product was redispersed in 20 mL of toluene for storage in a glass scintillation vial. The NC concentration was determined by weighing the dry pellet prior to redispersion. Typically, the pellet of $Fe_3O_4$ NCs redisperses fully.

**Size selection:** To further improve sample monodispersity, size selective precipitation was used. The $Fe_3O_4$@OA NC dispersion was brought to a concentration of ~15 mg/mL in toluene. The NC dispersion was added to a centrifuge tube that was being gently stirred. Ethanol was slowly added dropwise until the dispersion started turning cloudy. For 5 mL of sample this usually occurred when a total of ~1.5 mL of ethanol was added. The sample was centrifuged at 8000 g for 5 minutes. The largest NCs of the dispersion were destabilized and crashed out in this first step, leaving the smaller NCs dispersed. The NC dispersion was added to a clean centrifuge tube, while the precipitated NCs were dried under vacuum and redispersed in toluene. This process was repeated on the remaining NC solution with increasing ethanol content until all NC material had been extracted. The first size selection step contained the largest NCs, and the last step the smallest. Inspection of each size selection separation was performed by TEM, and samples of the same or similar sizes may be recombined. The largest and smallest NCs of the original size distribution were removed, yielding a more monodisperse sample.

**FICO NCs:** Fluorine and indium co-doped cadmium oxide NCs (FICO) were synthesized by following the procedure reported by Ye *et al.*[6] To preserve spherical shape, the dopant concentration was kept at 5%. To obtain the smallest sizes for this NC family, the metal to ligand molar ratio was kept at 1:3.5.

**Materials:** Cadmium acetylacetonate ($Cd(acac)_2$, ≥99.9% trace metals basis, Sigma Aldrich), Indium (III) fluoride ($InF_3$, ≥99.9% trace metals basis, Sigma Aldrich), oleic acid (OA, technical grade, Sigma Aldrich), 1-Octadecene (ODE, technical grade, Acros Organics).

**NC synthesis:** A 100 mL three-neck round-bottom flask was charged with 10 mg of $InF_3$, 354 mg of $Cd(acac)_2$, 1.2 g of OA, and 40 mL of ODE. The mixture was degassed while stirring at room temperature for 30 minutes. The atmosphere of the flask was cycled thrice between vacuum and nitrogen. The temperature was then increased to 75 °C, then to 125 °C, and the mixture was degassed for 1 hour. The flask was then filled with nitrogen and the temperature was quickly ramped to reach reflux, ~ 316 °C. Reflux was generally obtained within 6-8 minutes. After 10-20 minutes, the mixture abruptly changed in color to brown-green-orange. The temperature was maintained for an additional 15 minutes. The reaction was then quenched by removing the heating mantle and cooled with an air gun to 100 °C. The reaction was further cooled to room temperature by submerging the flask in a water bath. The NCs were separated from the reaction mixture by adding an equivalent volume of methyl acetate and centrifuging at 3500 g for 30 minutes. The precipitate was redispersed in hexanes and allowed to sediment overnight in a capped vial. The supernatant was then removed without disturbing the precipitate by using a syringe or pipette and filtered using a 200 nm PVDF syringe filter. The NCs were precipitated by adding an equivalent volume of isopropanol and centrifuging as described previously. This step was repeated once more, and the NCs were finally redispersed in 10 mL of toluene. The NC concentration was determined by drying a sacrificial fraction of the sample under vacuum overnight and weighing. Typically, the pellet of FICO NCs does not fully redisperse if allowed to dry fully.

## CHARACTERIZATION TECHNIQUES

*Ex situ* **small-angle X-ray scattering**:

The static patterns were collected using a Pilatus 1M detector on a Xeuss 2.0 system (Xenocs).

25 µL of a 10 mg/mL dispersion of NCs in toluene were loaded in a 1 mm capillary tube (Charles Supper). The capillary was then sealed by using a hot-glue gun.

The integration time for each measurement was set to 30 minutes. The sample to detector distance was set to 1.2 m. The beam energy was set to 8 keV (copper anode). The two-dimensional patterns were azimuthally averaged, and background subtracted. The intensity was calibrated to absolute units by using a capillary filled with water,[7] while the *q*-range was calibrated against a silver behenate standard. The scattering patterns were fitted to the form factor of a sphere

$$F(q) = 3 \frac{\sin qr - qr \cos qr}{(qr)^3}$$

convoluted with a Gaussian distribution of sizes to account for polydispersity by using the open-source SASfit software.[8]

*In situ* **small-angle X-ray scattering:**

The kinetic patterns were collected at the SMI beamline, Brookhaven National Laboratory by using a recently developed experimental setup,[9] expanded to support the simultaneous measurement of 4 samples by translating the sample stage vertically, Figure S4.

Each sample was prepared as follows: A 20 mL scintillation vial was charged with 8 mL of 200 mM sodium dodecyl sulfate in water. Subsequently, the vial was charged with 2 mL of a NC dispersion in 22v/v% toluene and 78v/v% hexanes with a total NC volume fraction of 0.001. The vial was capped and vigorously vortexed for 60 seconds using a vortex mixer (Fisher) to generate the emulsion.

The emulsion obtained is relatively polydisperse as shown by the results of Figure S10. We do not expect this to affect the assembly results since the scattered intensity of the X-rays is proportional to the number of unit cells – and therefore the volume - of the crystal. For instance, if we consider 2 droplets differing in radius by a factor of 2, then their volume and scattered intensity differ by a factor of 8. Therefore, the signal will be dominated by the largest droplets in the ensemble.

The emulsion was then uncapped and diluted by adding an additional 10 mL of 200 mM sodium dodecyl sulfate in water. A 1-inch octagonal stir bar was then added to the diluted emulsion. The vial was then placed on a hotplate (IKA plate) equipped with a thermocouple and a heating block for vials, heated to 70 °C while stirring at 500 rpm, and allowed to flow by means of a peristaltic pump (Cole-Palmer) at flow rate of 10 mL/minute through a closed loop of Viton peristaltic tubing (Cole-Palmer). The closed loop included a custom-made flow cell consisting of a 1 mm quartz capillary tube. The X-ray beam was aligned with the center of the capillary. This setup allowed us to measure the scattering pattern from the emulsion as evaporation occurred from the uncapped vial.

The integration time for each measurement was set to 1 second, the beam energy was 16.1 keV, and the sample to detector distance was 6.3 meters. The $q$-range was calibrated against a silver behenate standard. The two-dimensional patterns were azimuthally averaged, and background subtracted to yield $I(q,t)$, where $t$ is the time. The kinetic structure factor, $S(q,t)$, was then obtained by calculating $S(q,t) = I(q,t)/I(q,0)$ since at the beginning of the experiment the NCs are well dispersed within the droplets.

**Notes on AlB$_2$:** AlB$_2$ belongs to the family of binary crystal structures with hexagonal symmetry, characterized by a lattice parameter, $a$, and by the distance between the basal planes, $c$, with in general $a \neq c$. This structure is characterized by stacked hexagonal layers of the larger (L) NCs intercalated by hexagonal layers of the smaller (S) NCs occupying the trigonal prismatic voids left by the L NCs, see Figure S5.

The expected reflections $q_{hkl}$ for the planes of indexes $hkl$ are:

$$q_{hkl} = \sqrt{\frac{4}{3}\frac{h^2 + hk + k^2}{a^2} + \frac{l^2}{c^2}}$$

To proceed with the structural analysis, we first extracted the $c/a$ factor for the experimental AlB$_2$ binary NC superlattices (BNSLs) at the last time point, $t = t_{end}$. We did so empirically by generating the expected positions $q_{hkl}$ of the reflections for a given value of $c/a$, and comparing with the experimental results until an agreement was found. While the atomic AlB$_2$ structure shown in Figure S5 features a $c/a$ value of 1.084, the AlB$_2$ BNSLs we have investigated always showed values between 0.99 and 1.01, leading to an overall more symmetric structure. We also found that the value of $c/a$ does not vary as the structure evolves during drying.

To calculate the kinetic structural parameters, we first fitted each $S(q,t)$ curve with a superposition of Lorentzian curves with line shape:

$$L(q) = \frac{Aw^2}{(q - q_{hkl})^2 + w^2}$$

centered around the expected $q_{hkl}$ positions, with amplitude $A$ and full width at half maximum $2w$.

Since $q_{101}$ was the most isolated reflection, we chose to use it to calculate the kinetic structural parameters. From the Scherrer equation, we calculated the average crystal size as:

$$\xi = \frac{2\pi K}{2w}$$

where $K = 1.0747$ is the Scherrer constant used for a spherical crystal.[10]

For the superstructures isostructural to AlB$_2$, the lattice parameter, also equal to the center-to-center distance, or bond length, between L NCs was calculated as:

$$b_{LL} = a = \frac{2\pi}{q_{101}}\sqrt{\frac{4}{3} + \frac{1}{(c/a)^2}}$$

The surface-to-surface distance between A NCs was then calculated as:

$$d_{LL} = b_{LL} - \sigma_L$$

where $\sigma_L$ is the average diameter of the L NCs as measured by *ex situ* SAXS.

The bond length between S NCs was calculated by scaling by the expected values for the bond lengths in the atomic AlB$_2$ structure for which $b_{LL,At} = 0.30090\ nm$ and $b_{SS,At} = 0.17372\ nm$ so that:

$$b_{SS} = b_{LL}\frac{b_{SS,At}}{b_{LL,At}}$$

The surface-to-surface distance between S NCs was then calculated as:

$$d_{SS} = b_{SS} - \sigma_S$$

where $\sigma_S$ is the average diameter of the S NCs as measured by *ex situ* SAXS.

The bond length between L and S NCs was then calculated as:

$$b_{LS} = b_{LL}\frac{b_{LS,At}}{b_{LL,At}}$$

where $b_{LS,At}$ is the bond length in the atomic AB$_2$ structure for measured value of $c/a$ in the BNSL. For instance, for $c/a = 0.99$, $b_{AB,At} = 0.22883\ nm$. The surface-to-surface distance between L and S NCs was then determined as:

$$d_{LS} = b_{LS} - \sigma_L/2 - \sigma_S/2$$

Given its shape, the volume of the unit cell was calculated from:

$$V = a^2 c\sqrt{3}/2$$

Since each unit cell contains 1 L NC and 2 S NCs, the inorganic volume fraction of the NCs was determined as:

$$\phi = \frac{4\pi}{3V}[(\sigma_L/2)^3 + 2(\sigma_S/2)^3]$$

We also estimated the effective volume fraction of the NCs by taking in account the contribution of the ligands. To do so, we estimated the effective NC diameters as:

$$\sigma_{L,eff} = \sigma_L + min[d_{LL}(t = t_{end}), d_{LS}(t = t_{end})]$$
$$\sigma_{S,eff} = \sigma_S + min[d_{SS}(t = t_{end}), d_{LS}(t = t_{end})]$$

where $min(x, y)$ indicates the minimum value between $x$ and $y$. We then calculated the effective volume fraction of the NCs as:

$$\phi_{eff} = \frac{4\pi}{3V}\left[\left(\sigma_{L,eff}/2\right)^3 + 2\left(\sigma_{S,eff}/2\right)^3\right]$$

**Notes on NaZn$_{13}$:** NaZn$_{13}$ belongs to the family of binary crystal structures with cubic symmetry, characterized by a lattice parameter, $a$. It is useful to identify a simple-cubic sub cell of the structure of side length $a/2$ with 8 L NCs occupying the corners of the cube. The cube contains 1 body-centered S NC surrounded by an icosahedral cluster of 12 S NCs. The unit cell of the crystal structure is a cube of side length $a$ centered around this sub cell, see figure S13.

The expected reflections $q_{hkl}$ for the planes of indexes $hkl$ are:

$$q_{hkl} = \frac{2\pi}{a}\sqrt{h^2 + k^2 + l^2}$$

To calculate the kinetic structural parameters, we first fitted each $S(q, t)$ curve with a superposition of Lorentzian curves with line shape:

$$L(q) = \frac{Aw^2}{(q - q_{hkl})^2 + w^2}$$

centered around the expected $q_{hkl}$ positions, with amplitude $A$ and full width at half maximum $2w$.

Since $q_{200}$ was the most isolated reflection, we chose to use it to calculate the kinetic structural parameters. From the Scherrer equation, we calculated the average crystal size as:

$$\xi = \frac{2\pi K}{2w}$$

where $K = 1.0747$ is the Scherrer constant used for a spherical crystal.[10]

For the BNSLs isostructural to NaZn$_{13}$, the lattice parameter $a$, twice the center-to-center distance, or bond length, between L NCs $b_{LL}$, was calculated as:

$$2b_{LL} = a = \frac{4\pi}{q_{200}}$$

The surface-to-surface distance between L NCs was then calculated as:

$$d_{LL} = b_{LL} - \sigma_L$$

where $\sigma_L$ is the average diameter of the L NCs as measured by *ex situ* SAXS.

The bond length between S NCs was calculated by scaling by the expected values for the bond lengths in the atomic NaZn$_{13}$ structure for which $b_{LL,At} = 0.61365\ nm$ and $b_{SS,At} = 0.25664\ nm$ so that:

$$b_{SS} = b_{LL} \frac{b_{SS,At}}{b_{LL,At}}$$

The surface-to-surface distance between S NCs was then calculated as:

$$d_{SS} = b_{SS} - \sigma_S$$

where $\sigma_S$ is the average diameter of the S NCs as measured by *ex situ* SAXS.

The bond length between L and S NCs was then calculated as:

$$b_{LS} = b_{LL} \frac{b_{LS,At}}{b_{LL,At}}$$

where $b_{LS,At} = 0.35647\ nm$ is the bond length in the atomic NaZn$_{13}$ structure.

The surface-to-surface distance between L and S NCs was then determined as:

$$d_{LS} = b_{LS} - \sigma_L/2 - \sigma_S/2$$

Given its shape, the volume of the unit cell was calculated from:

$$V = a^3$$

Since each unit cell contains 8 L NCs and 104 S NCs, the inorganic volume fraction of the NCs was determined as:

$$\phi = \frac{4\pi}{3V}[8(\sigma_L/2)^3 + 104(\sigma_S/2)^3]$$

We also estimated the effective volume fraction of the NCs by taking in account the contribution of the ligands. To do so, we estimated the effective NC diameters as:

$$\sigma_{L,eff} = \sigma_L + min[d_{LL}(t = t_{end}), d_{LS}(t = t_{end})]$$
$$\sigma_{S,eff} = \sigma_S + min[d_{SS}(t = t_{end}), d_{LS}(t = t_{end})]$$

where $min(x, y)$ indicates the minimum value between $x$ and $y$. We then calculated the effective volume fraction of the NCs as:

$$\phi_{eff} = \frac{4\pi}{3V}\left[(\sigma_{A,eff}/2)^3 + 2(\sigma_{B,eff}/2)^3\right]$$

During the assembly with a 1:13 NC number ratio, we observed the coexistence of the majority $NaZn_{13}$ phase with a minority $AlB_2$ phase. To calculate the fraction of the minority phase, $f_{AlB2}$, we took the ratio of the time-dependent area of the 101 reflection, $Area_{101} = \pi A_{101} w_{101}$, measured at a NC number ratio of 1:13, with the late-time area of the 101 reflection measured for the same NCs at a number ratio of 1:2. That is:

$$f_{AlB2,1:13}(t) = \frac{Area_{101,1:13}(t)}{Area_{101,1:2}(t = t_{end})}$$

We calculated the fraction of the majority phase, $f_{NaZn13}$, by using:

$$f_{NaZn13,1:13}(t) = \frac{Area_{200,1:13(t)}}{max[Area_{200,1:13}(t)]} \times \{1 - max[f_{AlB2,1:13}(t)]\}$$

**Transmission electron microscopy**: For low-resolution TEM, a JEOL 1400 microscope was operated at 120 kV. For high-resolution TEM and STEM, a JEOL F200 microscope was operated at 200 kV. During imaging, magnification, focus and tilt angle were varied to yield information about the structure of the BNSLs.

To prepare the dispersed NCs for imaging, we drop casted 10 μL of a dilute (~0.1 mg/mL) dispersion of NCs in toluene on a carbon-coated TEM grid (EMS). The grid was dried under vacuum for 1 hour prior to imaging. To determine the effective size of the NCs, we first imaged an area containing a well-ordered monolayer of NCs. We then opened the micrograph using ImageJ and performed a fast-Fourier transformation (FFT). Hovering with the cursor over the spots in the FFT yielded the periodicity of the monolayer. A circle was fit to the six spots in the FFT image and the effective size of the NCs was determined to be equal to the period corresponding to the radius of the circle, see Figure S2.

To prepare the superstructures for imaging, after the emulsion has fully dried, the binary NC superstructures were washed twice in 20 mM sodium dodecyl sulfate in water by centrifugation at 3000 g for 30 minutes, and redispersed. 10 μL of the dispersion was drop casted on a carbon-coated TEM grid (EMS) and dried under vacuum for 1 hour. The grid was then dipped in a cleaning solution consisting of 1:2 water:isopropranol, and dried for 1 hour under vacuum.

**Scanning electron microscopy:** SEM characterization was conducted using a JEOL 7500F SEM. Images were acquired using the high-resolution setting, a beam energy of 30 kV, a 20 μA beam current, a probe current setting of 10, and a WD = 4.3 mm using the SEI detector. Energy dispersive spectroscopy (EDS) and mapping were performed using an EDAX APEX system with a silicon drift detector (SDD) using an accelerating voltage of 30 kV and a live time of 2949 s.

**Magnetic measurements:** DC magnetic measurements were collected using a Quantum Design Brand Physical Property Measurement System (PPMS) Superconducting Quantum Interference Design (SQUID) Magnetometer. Field and Zero Field Cooling curves were taken by stepping the temperature between 15K and room temperature, and hysteresis curves were obtained between -3T and 3T at 15K and 300K respectively. Cooling was performed using liquid helium. All measurements were performed under high vacuum. For each measurement, approximately 50uL of sample was placed into the sample holder and inserted into the SQUID using a nonmagnetic sample rod.

**Spectrophotometry**: Absorption spectra of NC dispersions in tetrachloroethylene were measured by using a Cary 5000 UV-Vis-NIR spectrophotometer, Figure S3.

## SUPPORTING REFERENCES